\documentclass[traditabstract]{aa}
\usepackage{epsfig,caption}
\usepackage{multirow}
\usepackage{graphics}
\usepackage[latin1]{inputenc}
\usepackage{natbib}

\usepackage[squaren, Gray, cdot]{SIunits}

\begin{document}

\title{\bf The one-dimensional Ly$\alpha$ forest power spectrum from BOSS }
\author{Nathalie Palanque-Delabrouille\inst{1} \and Christophe Y\`eche\inst{1} \and Arnaud Borde\inst{1}  \and Jean-Marc Le Goff\inst{1}, Graziano Rossi\inst{1},  Matteo Viel \inst{2,3},
\'{E}ric Aubourg\inst{4},
Stephen~Bailey\inst{5},
Julian~Bautista\inst{4},
Michael~Blomqvist\inst{9},
Adam Bolton\inst{7},
James S.~Bolton\inst{19},
Nicol\'as G. Busca\inst{4}, 
Bill Carithers\inst{5},
Rupert A.C. Croft\inst{6},
Kyle S. Dawson\inst{7},
Timoth\'ee~Delubac\inst{1}, 
Andreu Font-Ribera\inst{5,8},
Shirley Ho\inst{6},
David~Kirkby\inst{9},
Khee-Gan Lee\inst{10},
Daniel~Margala\inst{9},
Jordi~Miralda-Escud\'{e}\inst{11,12},
Demitri Muna\inst{22},
Adam D. Myers\inst{13},
Pasquier Noterdaeme\inst{14},
Isabelle P\^aris\inst{14,15},
Patrick Petitjean\inst{14},
Matthew M. Pieri\inst{16},
James Rich\inst{1},
Emmanuel Rollinde\inst{14},
Nicholas P. Ross\inst{5},
David~J.~Schlegel\inst{5},
Donald~P.~Schneider\inst{20,21},
An\v{z}e Slosar\inst{17}
\and David H. Weinberg\inst{18}} 
\institute{CEA, Centre de Saclay, Irfu/SPP,  F-91191 Gif-sur-Yvette, France
\and
INAF, Osservatorio Astronomico di Trieste, Via G. B. Tiepolo 11, 34131 Trieste, Italy
\and
INFN/National Institute for Nuclear Physics, Via Valerio 2, I-34127 Trieste, Italy.
\and 
APC, Universit\'{e} Paris Diderot-Paris 7, CNRS/IN2P3, CEA,
       Observatoire de Paris, 10, rue A. Domon \& L. Duquet,  Paris, France
\and
Lawrence Berkeley National Laboratory, 1 Cyclotron Road, Berkeley, CA 94720, USA
\and
Bruce and Astrid McWilliams Center for Cosmology, Carnegie Mellon University, Pittsburgh, PA 15213, USA
\and
Department of Physics and Astronomy, University of Utah, 115 S 1400 E, Salt Lake City, UT 84112, USA
 \and
Institute of Theoretical Physics, University of Zurich, 8057 Zurich, Switzerland 
  \and
Department of Physics and Astronomy, University of California, Irvine, CA 92697, USA
\and
Max-Planck-Institut f\"ur Astronomie, K\"onigstuhl 17, D69117 Heidelberg, Germany
\and
Instituci\'{o} Catalana de Recerca i Estudis  Avan\c{c}ats, Barcelona, Catalonia
\and
Institut de Ci\`{e}ncies del Cosmos, Universitat de Barcelona/IEEC, Barcelona 08028, Catalonia
\and
Department of Physics and Astronomy, University of Wyoming, Laramie, WY 82071, USA
\and
Universit\'e Paris 6 et CNRS, Institut d'Astrophysique de Paris, 98bis blvd. Arago, 75014 Paris, France
\and
Departamento de Astronom\'ia, Universidad de Chile, Casilla 36-D, Santiago, Chile
\and
Institute of Cosmology and Gravitation, Dennis Sciama Building, University of Portsmouth, Portsmouth, PO1 3FX, UK
 \and
Bldg 510 Brookhaven National Laboratory, Upton, NY 11973, USA
\and
Department of Physics and Center for Cosmology and Astro-Particle Physics, Ohio State University, Columbus, OH 43210, USA
\and
School of Physics and Astronomy, University of Nottingham,  University Park, Nottingham, NG7 2RD, UK
\and
Department of Astronomy and Astrophysics, The Pennsylvania State University, University Park, PA 16802, USA
\and
Institute for Gravitation and the Cosmos, The Pennsylvania State University, University Park, PA 16802, USA
\and
Department of Astronomy, Ohio State University, Columbus, OH, 43210, USA
%  \and
%Apache Point Observatory, P.O. Box 59, Sunspot, NM 88349, USA
%\and
%Institute for Advanced Study, Einstein Drive, Princeton, NJ 08540, USA
%\and
%Harvard-Smithsonian Center for Astrophysics, Harvard University, 60 Garden St., Cambridge MA 02138, USA
%\and
%Center for Cosmology and Particle Physics, New York University, New York, NY 10003, USA
%\and
%Department of Astronomy, University of Wisconsin, 475 North Charter Street, Madison, WI 53706, USA
%\and
%Steward Observatory, University of Arizona, 933 N. Cherry Ave., Tucson, AZ 85121, USA
%\and
%INAF, Osservatorio Astronomico di Trieste, Via G. B. Tiepolo 11, 34131 Trieste, Italy
%\and
%INFN/National Institute for Nuclear Physics, Via Valerio 2, I-34127 Trieste, Italy.
%\and
%Department of Astronomy, Ohio State University, 140 West 18th Avenue, Columbus, OH 43210, USA
%\and
%Department of Astronomy and Astrophysics and the Enrico Fermi Institute, The University of Chicago, 5640 South Ellis Avenue, Chicago, Illinois, 60615, USA
}
\date{Received xx; accepted xx}
\authorrunning{Palanque-Delabrouille, N.  et al.}
\titlerunning{1D Lyman-$\alpha$ power spectrum from BOSS}
\abstract{We have developed two independent methods for measuring the one-dimensional power spectrum of the transmitted flux in the Lyman-$\alpha$  forest. The first method is based on a Fourier transform and  the second on a maximum-likelihood estimator. The two methods are independent and have different systematic uncertainties.   Determination of the noise level in the data spectra was subject to a new  treatment, because of its significant impact  on the derived power spectrum.

We applied the two methods to 13~821 quasar spectra from SDSS-III/BOSS DR9  selected from a larger sample of over 60~000 spectra on the basis of their high quality,  high signal-to-noise ratio (S/N), and  good spectral resolution. The  power spectra  measured using either approach  are in good agreement over all twelve redshift bins from $\langle z\rangle = 2.2$ to $\langle z\rangle = 4.4$, and  scale from 0.001~$\rm(km/s)^{-1}$ to $0.02~\rm(km/s)^{-1}$. We determined the methodological and instrumental systematic uncertainties of our measurements.

We provide a preliminary cosmological interpretation of our measurements using available hydrodynamical simulations. The improvement in precision over  previously published results from SDSS is a factor 2--3 for constraints on relevant cosmological parameters. For a $\Lambda$CDM model and using a constraint on $H_0$ that encompasses  measurements based on the local distance ladder and on CMB anisotropies, we infer $\sigma_8 =0.83\pm0.03$ and $n_s= 0.97\pm0.02$ based on  \ion{H}{i} absorption in the range  $2.1<z<3.7$. \thanks{The  
measured values of the power spectrum and correlation matrices for all scales and all redshifts, corresponding to Tables~\ref{tab:Resultsfft} and \ref{tab:Resultslkl}, are only available in electronic form at the CDS via anonymous ftp to cdsarc.u-strasbg.fr (130.79.128.5), or via http://cdsweb.u-strasbg.fr/cgi-bin/qcat?J/A+A/.} }

\maketitle

\section{Introduction}
  Neutral hydrogen in the intergalactic medium scatters light at the 
Lyman-$\alpha$ absorption wavelength
$\lambda_{\rm Ly\alpha}\sim1216\,$\AA, producing an absorption spectrum
that is observed on any background source as a map of transmission
fraction as a function of redshift~\citep{bib:lynds71}. At high redshift, when the typical
absorption from intergalactic matter is  sufficiently strong, the continuous
nature of the absorption spectrum is easily observable as the
Lyman-$\alpha$ (or Ly$\alpha$) forest. Even though this spectrum may be fitted
as a series of merged absorption lines, simulations reveal that it is in reality a map of the
density fluctuations in the intervening intergalactic medium
seen in redshift space, with peaks of absorption at the density peaks of
the absorbing gas \citep{bib:bi92, bib:miraldaescude93}. 
In fact, the fluctuations in the Ly$\alpha$ forest absorption can be
used as a tracer of the varying density of intergalactic gas expected
from the growth of structure from primordial fluctuations in the
Universe~\citep{bib:croft98}.
The physics at play is  understood well for an  intergalactic
medium that is heated exclusively by photoionization,
and it can be modeled with hydrodynamic simulations~\citep{bib:cen94, bib:zhang95, bib:hernquist96, bib:hui97a, bib:hui97b}, although additional heating mechanisms,
such as radiative transfer effects during 
hydrogen and helium reionization~\citep{bib:abel99}, and the complex mechanical
effects of galactic winds and quasar outflows may modify this simple
picture.

The information embedded in the Ly$\alpha$ forest can  be
used to probe the amplitude and shape of the power spectrum
of mass fluctuations~\citep{bib:croft98, bib:gnedin98, bib:hui99,
bib:gaztanaga99, bib:nusser99, bib:feng00, bib:mcdonald00,
bib:hui01} and to constrain cosmology through the study of redshift-space distortions and the Alcock-Paczynski test~\citep{bib:AP79, bib:hui99, bib:mcdonald99, bib:croft02}, 
the mass of neutrinos~\citep{bib:seljak05, bib:viel10}, 
or the BAO peak position~\citep{bib:mcdonald07}.
Initially, the Ly$\alpha$  forest power spectrum was studied exclusively
along the line of sight by measuring the correlation separately in
each quasar spectrum, starting with
the use of small numbers of high-resolution spectra: 1 Keck HIRES
spectrum \citep{bib:croft98}, 19 spectra from the Hershel telescope
on La Palma or the AAT~\citep{bib:croft99}, 8 Keck HIRES spectra \citep{bib:mcdonald00}, a set of 30 Keck HIRES and 23 Keck LRIS spectra~\citep{bib:croft02}, or a set of 27 high-resolution UVES/VLT QSO
spectra at redshifts $\sim$ 2 to 3~\citep{bib:kim04a, bib:kim04b,
bib:viel04}. 

A substantial breakthrough was achieved with the
measurement of the Ly$\alpha$ forest power spectrum based on the much
larger sample of 3035 medium-resolution ($R=\Delta \lambda/\lambda \approx 2000$) quasar spectra from the Sloan
Digital Sky Survey \citep{bib:york00} by~\cite{bib:mcdonald06}. The large number of
observed quasars allowed detailed measurements with well characterized
errors of the power spectrum up to larger scales, probing the linear regime and providing cosmological
constraints~\citep{bib:mcdonald05b, bib:seljak05}.

Recently, the  Sloan Digital Sky Survey III
\citep{bib:eisenstein11} has carried out the Baryon Oscillation
Spectroscopic Survey \citep{bib:dawson13}. This new survey
has been especially designed to target quasars at redshift $z>2$, which
are useful for the Ly$\alpha$ forest analysis and to obtain spectra of
many more of them than in the previous phases of SDSS
(see \citet{bib:dawson13} and references therein). This large number of quasars
allowed for a detailed measurement of the Ly$\alpha$ power spectrum in
3D redshift space (as a function of the transverse and
parallel directions) in \citet{bib:slosar11}, using the first 14000
quasars of the BOSS survey. For the first time, the redshift distortions
predicted in linear theory of large-scale structure by gravitational
evolution \citep{bib:kaiser87} were detected in the Ly$\alpha$ forest. This is
in fact the highest redshift detection of redshift distortions that has
been achieved in observational cosmology with any large-scale structure
tracer. With the quasars in the Data
Release 9 \citep{bib:ahn12}, containing more than 60000 quasars with observed Ly$\alpha$
forest absorption \citep{bib:paris12, bib:KG13}, the measurement of the redshift
space power spectrum has been extended up to the scales of the Baryon
acoustic oscillations (hereafter BAO), yielding the highest redshift
measurement of the BAO peak position and
providing new constraints on the history of the expansion of the
universe \citep{bib:busca13, bib:slosar13, bib:kirkby13}.

The measurement of the 3D power spectrum uses only
information from the flux correlation of pixel pairs in different
quasar spectra that are relatively close in the sky. However, the
correlation of pixel pairs on the same quasar spectrum provides
complementary, useful information on the Ly$\alpha$ correlation along
the line of sight, which is also important for constraining the physical
parameters of the Ly$\alpha$ forest. The 1D power
spectrum, $P_{1D}(k_\parallel)$ (equal to the 1D Fourier transform (hereafter FT) of the
correlation function along the line of sight), is related to the
3D one by
\begin{equation}
  P_{1D}(k_\parallel) = \int_0^\infty {dk_\perp k_\perp\over 2\pi}\,
 P_{3D}(k_\parallel,k_\perp) ~. 
\label{eq:P3DvsP1D}
\end{equation}
If all the relevant scales could be treated with in the limit of linear
theory, the 3D power spectrum should be simply related to
the mass power spectrum according to
$P_{3D}(k_\parallel,k_\perp) =
b_\delta^2 P(k) (1+\beta k_\parallel^2/k^2)^2$, where
$k^2 = k_\parallel^2 + k_\perp^2$, and $b_\delta$ and $\beta$ are the
density bias and redshift distortion parameters of the Ly$\alpha$
forest \citep{bib:mcdonald03, bib:slosar11}. However, linear theory is valid only on large scales, and even though the linear expression is valid for $P_{3D}$ when k is small, the 1D
$P_{1D}$ is affected by the non-linearities of small scales even for very
low values of $k_\parallel$ in Eq.~\ref{eq:P3DvsP1D}.  The theoretical interpretation of
measurements of $P_{1D}$ is therefore always dependent on the nonlinear
physics of the intergalactic medium on small scales.

 In the present paper, we measure the 1D transmission power spectrum of
the Ly$\alpha$ forest from a sample of
13,821  quasar spectra, which are selected as the highest
quality spectra among the set of
61931 quasars at $z>2.15$ from the DR9 quasar catalog of
\citet{bib:paris12}.

  Historically, two approaches have been used to measure the 1D power
spectrum of the fluctuations in the transmitted flux fraction $F$. The first
is done directly in Fourier space by computing the FT of
$\delta = F/\langle F\rangle - 1$ for each quasar spectrum and obtaining
the power spectrum from these Fourier modes, as
in~\citet{bib:croft98, bib:croft02} and \citet{bib:viel04}. The second approach
uses a likelihood method to compute  the covariance matrix
of $\delta$ in real space (or line-of-sight correlation function) as a function of the
pixel pair separation in the spectra~\citep{bib:mcdonald06}. The 1D power
spectrum is the FT of the Ly$\alpha$ correlation function
obtained in this way. The two methods have their own advantages and
drawbacks in terms of, for example, robustness, processing speed, accounting of instrumental effects, precision, etc. To benefit from their complementarity, we have developed independent analysis pipelines based on either technique. In this paper we present and compare  the results
obtained with the two approaches. 

The outline of our paper is as follows. In section~\ref{sec:BOSSdata},
we present the BOSS data and explain how we calibrate the level of noise in
the spectra and determine the spectrograph resolution.  The selection of the quasar spectra  and the different steps of the data preparation  are presented in section~\ref{sec:data}. 
In section~\ref{sec:methods}, we describe the two complementary methods we have
developed to analyze the data. We present in
section~\ref{sec:systs} our estimates of the systematic uncertainties
associated with each method or due to our imperfect knowledge of the instrument performances. The final results are given in
section~\ref{sec:results}, and a preliminary cosmological interpretation 
is presented in section~\ref{sec:cosmo}, along with a comparison
to previously published constraints. Conclusions and perspectives are
presented in section~\ref{sec:discussion}.

%%%%%%

%\section{Data preparation }\label{sec:BOSSdata}
\section{Data calibration}\label{sec:BOSSdata}

\subsection{BOSS survey}
 \label{sec:dataprep}
The Sloan Digital Sky Survey \citep{bib:york00} mapped over one
quarter of the sky using the dedicated 2.5-m Sloan Telescope
\citep{bib:gunn06} located at Apache Point Observatory in New Mexico.  A
 mosaic CCD camera \citep{bib:gunn98} used in drift-scanning mode imaged this area in five
photometric bandpasses \citep{bib:fukugita96,bib:smith02, bib:doi10} to a limiting
magnitude of $g\simeq 22.8$.  The imaging data were processed through
a series of pipelines~\citep{bib:stoughton02} that performed astrometric calibration, photometric reduction, and photometric
calibration.  The magnitudes were corrected for Galactic
extinction using the maps of \citet{bib:schlegel98}. As part of the
SDSS-III survey \citep{bib:eisenstein11}, BOSS  imaged an additional $3~000$ square
degrees of sky over that of SDSS-II \citep{bib:abazajian09} in the southern
Galactic sky and in a manner identical to the original SDSS imaging.
This increased the total imaging SDSS footprint to $14~055$ square
degrees, with $7~600$ square degrees at Galactic latitude $|b| > 20$ deg in the northern
Galactic cap and $3~000$ square degrees at $|b| > 20$ deg in the southern
Galactic cap.  All of the imaging was reprocessed and released as
part of SDSS Data Release~8 \citep{bib:aihara11}.

BOSS is a spectroscopic survey primarily designed to obtain spectra and redshifts over a footprint covering 10~000 square degrees for 1.35 million galaxies, 160~000 quasars, and approximately 100~000 ancillary targets. The quasars, whose spectra cover the Ly$\alpha$ forest of interest for this work, are selected with several algorithms based on the SDSS imaging~\citep{bib:yeche10, bib:kirkpatrick11,  bib:Bovy11, bib:palanque11}, which are all summarized in \citet{bib:ross12}. A full description of the BOSS survey design is given in  \citet{bib:dawson13}. 
Aluminum plates are drilled with 1000 holes whose positions
correspond to the positions of the targets on the focal plane of the telescope. They are manually
plugged with optical fibers that feed a pair of double spectrographs.
The double-armed BOSS spectrographs are significantly upgraded from
those used by SDSS-I/II, covering the wavelength range $3~600\,$\AA\ to
$10~000\,$\AA\ with a resolving power of 1~500 to 2~600~\citep{bib:smee12}.
In addition to expanding the wavelength coverage relative to the previous $3850$--$9200\,$\AA\ range 
of SDSS-I, the throughputs have been increased with new CCDs,
gratings, and improved optical elements, and the 640-fiber cartridges
with $3''$ apertures have been replaced with 1~000-fiber cartridges with
$2''$ apertures. Each observation is performed in a series of
900-second exposures, integrating until a minimum S/N is achieved for the faint galaxy targets. 

% REFS SDSS
%http://adsabs.harvard.edu/abs/2012arXiv1208.2233S Smee et al
%http://adsabs.harvard.edu/abs/2012arXiv1208.0022D Dawson
%http://adsabs.harvard.edu/abs/2012arXiv1207.7326B Bolton
%http://adsabs.harvard.edu/abs/2012arXiv1207.7137S Ahn DR9      ---------
%http://adsabs.harvard.edu/abs/2011AJ....142...72E Eisenstein

\subsection{BOSS reduction pipeline}
The data are reduced using a pipeline adapted for BOSS from the SDSS-II spectroscopic reduction pipeline~\citep{bib:bolton12}. All the spectra are wavelength-calibrated, sky-subtracted, and flux-calibrated. 
The final spectrum for a given object is produced by the coaddition of typically four to seven 900-second individual exposures that can be distributed over several nights of observations.  The coadded spectrum is rebinned onto a uniform baseline of $\Delta \log_{10} (\lambda) = 10^{-4}$  per pixel. The pipeline  computes a statistical error estimate for each pixel, incorporating photon noise, CCD read-out noise, and sky-subtraction errors. 

For each spectrum, the pipeline also provides a spectral classification and a redshift for the extragalactic objects. A visual inspection is then performed on the spectra of all quasar targets to provide the final classification and redshifts~\citep{bib:paris12}. 

At low redshift ($z<2.5$), the 1D power spectrum has a significant contribution from photon noise, so it is quite sensitive to the precision with which the noise level in the data is known. The spectrograph wavelength resolution is also a major issue on small scales (i.e., large $k$-modes) where it abruptly reduces the power spectrum by  a factor of $\sim$2 at $k=0.01\;{\rm (km/s)}^{-1}$ and by a factor of 5--10 at $k=0.02\;{\rm (km/s)}^{-1}$. The accuracy with which noise and spectrograph resolution are determined in the automated pipeline is insufficient for the purpose of this analysis. We have therefore developed techniques to derive corrections, described in the following sections. These refinements were not necessary for  measuring  the large-scale 3D Ly$\alpha$ correlation function~\citep{bib:busca13,bib:slosar13} since the BAO feature occurs on much larger scales than the size of the resolution element, and the noise in the data only affects the amplitude of the power spectrum and  not  the correlation function where the BAO peak is seen. Instead, we here aim at measuring the absolute level of the power spectrum, which is directly affected by the level of noise, down to scales of  a few Mpc, i.e., of a few pixels, where an accurate knowledge of the spectrograph resolution is crucial.

\subsection{Calibration of pixel noise}
\label{sec:noise}

\begin{figure}[h]
\begin{center}
\epsfig{figure= 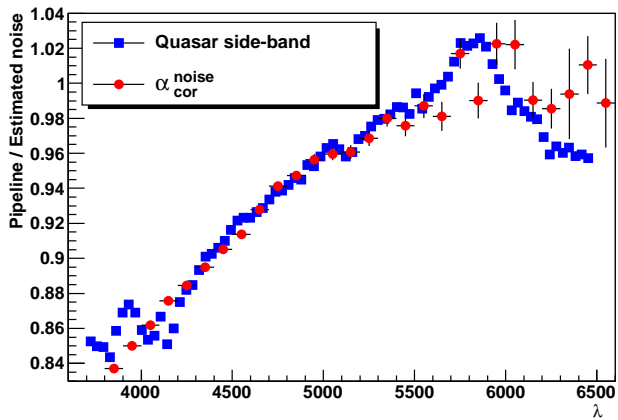, width = \columnwidth} 
\caption[]{\it Ratio of the pipeline noise estimate to the actual flux dispersion in the spectra. The blue squares denote this ratio as estimated from the quasar $1330<\lambda_{\rm RF}<1380\,$\AA\ and $1450<\lambda_{\rm RF}<1500\,$\AA\ sidebands. The red points indicate the correction from our procedure (Eq.~\ref{eq:corr}) as a function of mean forest wavelength.} 
\label{fig:noise}
\end{center}
\end{figure}

The noise  provided by the SDSS-III pipeline is known to suffer from systematic underestimates e.g., \citep{bib:mcdonald06, bib:desjacques07}. To investigate the extent of this issue, we examined the pixel variance in spectral regions that are intrinsically smooth and flat. We used two $50\,$\AA\ regions of quasar spectra (hereafter `side-bands'), redwards of the Ly$\alpha$ peak: $1330<\lambda_{\rm RF}<1380\,$\AA\  and $1450<\lambda_{\rm RF}<1500\,$\AA.  These bands are  not  affected by Ly$\alpha$ forest absorption and have a quasar unabsorbed flux that is relatively flat with  wavelength. For each individual quasar, we  computed the ratio of the mean pipeline error estimate in the band, $\langle \sigma_p\rangle$, to the root-mean-square (rms) of the pixel-to-pixel  flux dispersion  within the same band. This quantity is  averaged over all DR9 quasars, giving us a wavelength-dependent measure of the accuracy of the pipeline noise estimate because of the distribution of quasar redshifts (see Fig.~\ref{fig:noise}). For a perfect noise estimation, the plotted quantity should be unity at all wavelengths; on the other hand, under (over) estimates will produce values below (above) unity. The flux dispersion in the blue part of the spectra ($\lambda<4000\,$\AA) is seen to be about 15\% larger than expected from the noise level given by the pipeline. The discrepancy decreases with increasing wavelength, and the two estimates are in agreement at $\lambda\simeq 5700\,$\AA. 

This test clearly indicates a wavelength-dependent miscalibration of the noise.  However, 
since some of the flux dispersion in the quasar sidebands can arise from intervening metals along the sightline (see correction of the metal contribution to the power spectrum in Sec.~\ref{sec:metals}), this procedure could  overestimate the true noise.  In~\cite{bib:KG13}, we  provided a per-quasar correction to the pipeline noise that was sufficient for BAO studies, but still not accurate enough for this power-spectrum analysis. Here, we recalibrate the pixel noise for each quasar as described below. This new  correction  deviates from the one described in  \cite{bib:KG13} at most by a few percent.

We  made use of the typically four to seven individual exposures that contribute to a given quasar spectrum and split them into two interleaved sets: one containing  the odd  and the other the even exposures.  For each set, we computed the weighted average spectrum with weights equal to the pixel inverse variance  $\sigma_p^{-2}$ given by  the  pipeline of BOSS, binned into pixels of width $\Delta \log_{10}(\lambda) = 10^{-4}$ as for the final coadded spectrum. We then computed a `difference spectrum' $\Delta\phi$ by subtracting the spectrum obtained for one set from the one for the other set. In this process, we mask all pixels affected by sky emission lines (cf. Sec.~\ref{sec:methods_syst}) by setting to 0 the value of the corresponding pixel in the difference spectrum. The difference spectrum should have all physical signal removed and only contain signal fluctuations. It can therefore be used to directly determine the level of noise in the data, irrespective of any miscalibration of the pixel noise in the reduction pipeline. This procedure also has the advantage of evaluating the noise level for each individual spectrum and not on a statistical basis.

\begin{figure}[h]
\begin{center}
\epsfig{figure= 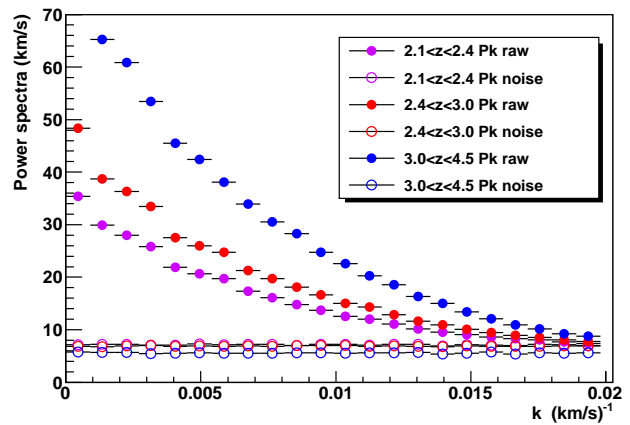,width = \columnwidth} 
\caption[]{\it Average power spectra of the raw (filled dots) and of the difference (open circles) signal for three ranges in Ly$\alpha$ redshifts. }
\label{fig:diff_pk}
\end{center}
\end{figure}

We computed the quantity $P_{\rm diff}^{noise} = | \mathcal{ F} (\Delta\phi) |^2$, where $\mathcal{ F} (\Delta\phi)$ is the FT of the difference spectrum $\Delta\phi$.
In Fig.~\ref{fig:diff_pk}, we plot the average of $P_{\rm diff}^{noise}$  computed over the Ly$\alpha$ forest of quasars, for three ranges in Ly$\alpha$ redshifts (or equivalently three ranges in observed wavelength). The noise is expected to be white, and $P_{\rm diff}^{noise}$  is indeed seen to be scale-independent to an accuracy sufficient for our purposes.  For comparison, we also show in the figure the power spectrum of  coadded spectra where both signal and noise are present. The noise power spectrum approaches the same order of magnitude as the raw power spectrum on small scales ($k\sim 0.02\;({\rm km/s})^{-1}$) and low redshifts ($z<2.4$). This is therefore the region where it is most important to accurately determine its contribution.

We derived the `pipeline noise power spectrum'  $P_{\rm pipe}^{noise}$ from the error $\sigma_p$ given by the pipeline in each pixel. $P_{\rm pipe}^{noise}$ would be the true noise power spectrum if the pipeline error estimate were correct.  For each individual quasar, we thus derive a correction coefficient of the pixel flux error as
\begin {equation}
\alpha_{\rm cor}^{noise} = \sqrt{\langle P_{\rm pipe}^{noise} \rangle / \langle P_{\rm diff} ^{noise}\rangle} \;\; ,
\label{eq:corr}
\end{equation}
where the power spectra are computed in both cases over the pixels in the quasar forest and  averaged  over $k$. 
In Fig.~\ref{fig:noise}, the value of the correction term is shown, averaged over all the DR9 quasars; as before,  the distribution of quasar redshifts provides a wavelength-dependent measurement. We observe, on average,  excellent agreement between $\alpha_{\rm cor}^{noise}$ and the noise miscalibration estimated in quasar sidebands. In the latter case, however,  the estimate is derived from lower redshift quasars whose sideband covers the same wavelength region as the Ly$\alpha$ forest of higher redshift quasars. The method based on spectrum differences, in contrast, uses the forest data directly and is thus a better estimate of the noise in each quasar spectrum. For each quasar, the corrected pixel error $\sigma$ is  derived from the pipeline pixel error $\sigma_p$ by $\sigma(\lambda) = \sigma_p(\lambda) / \alpha_{\rm cor}^{noise}$.

\subsection{Calibration of spectrograph resolution}
\label{sec:psf}

For each co-added spectrum, the spectral resolution is provided by the BOSS reduction pipeline~\citep{bib:bolton12}. Since the measurement of the 1D power spectrum on small scales is extremely sensitive to the spectrograph resolution, we  first investigated the resolution given by the pipeline and we  determined a correction table.

\subsubsection{Spectrograph resolution in the BOSS pipeline}
In BOSS, spectral lamps are used to provide the wavelength calibration, as described in~\citet{bib:smee12}. In the present work, we are mostly interested in the calibration of the blue CCD, which is obtained from its illumination with a mercury-cadmium arc lamp (with seven principal emission lines in the blue and the green parts of the spectrum). 

The spectral resolution is measured from calibration arc lamp images taken before each set of science exposures. The pipeline procedure fits a Gaussian distribution around the position of the mercury and cadmium lines. The mean $m_\lambda$ and the width $\sigma_{\lambda}$ of the Gaussian  determine the absolute wavelength on the CCD and the resolution of the spectrograph, respectively. A fourth-order Legendre polynomial is fit to the derived $\sigma_{\lambda}$ as a function of  wavelength to model the dispersion over the full wavelength range. 

\subsubsection{Precision of  pipeline resolution}

\begin{figure*}[htbp]
\begin{center}
\epsfig{figure= 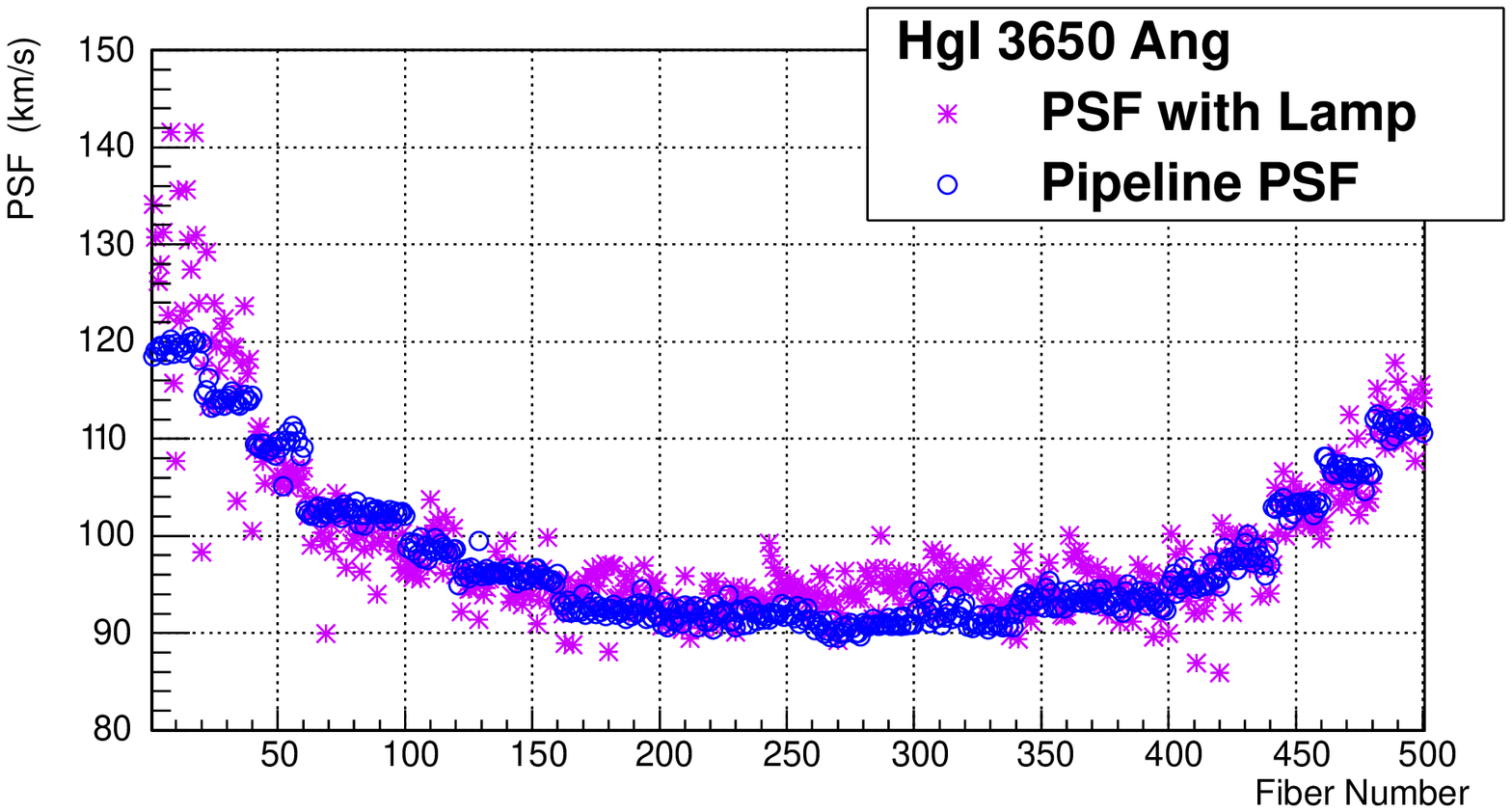, width = \columnwidth} \epsfig{figure= 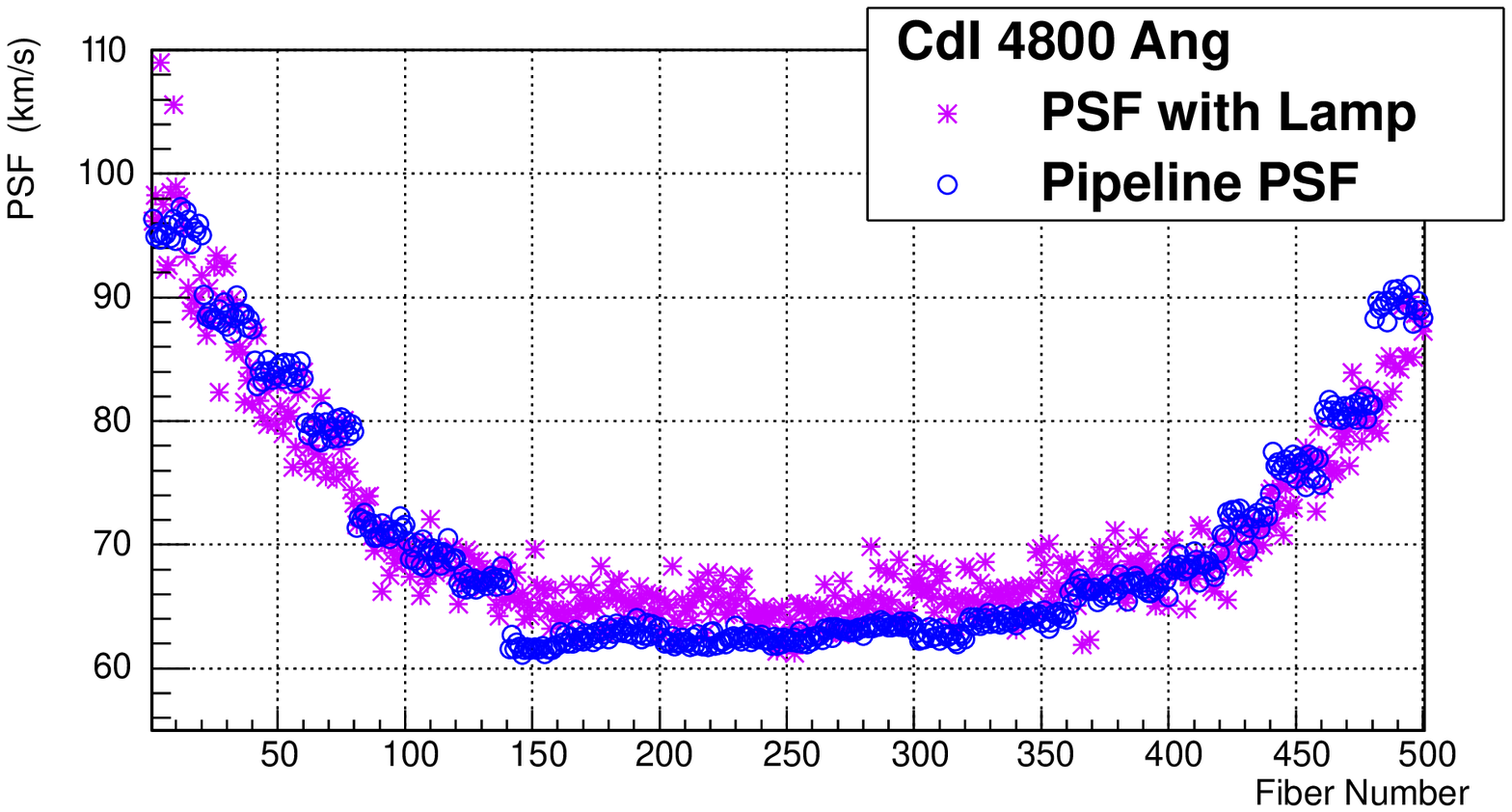, width = \columnwidth} 
\epsfig{figure= 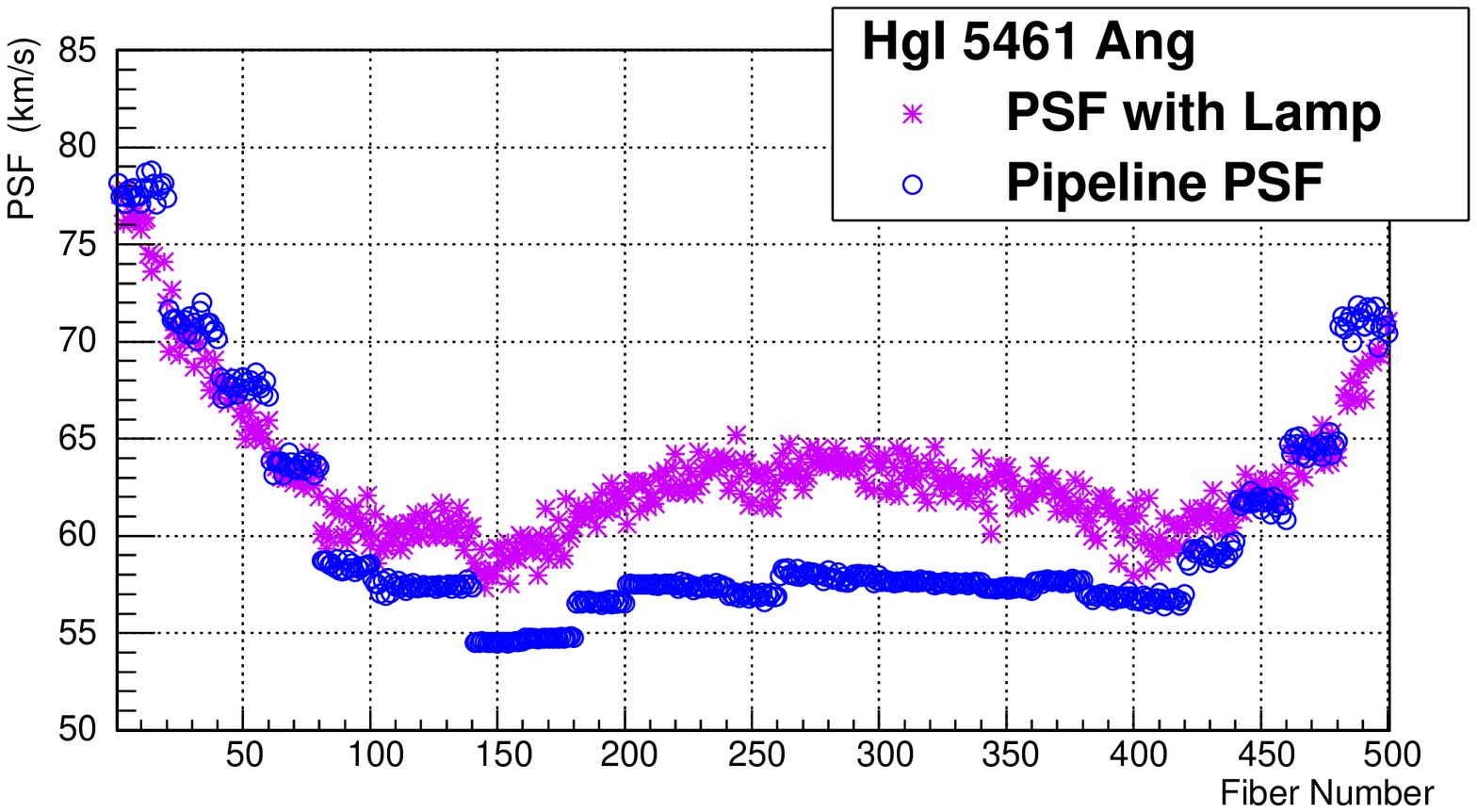, width = \columnwidth} \epsfig{figure= 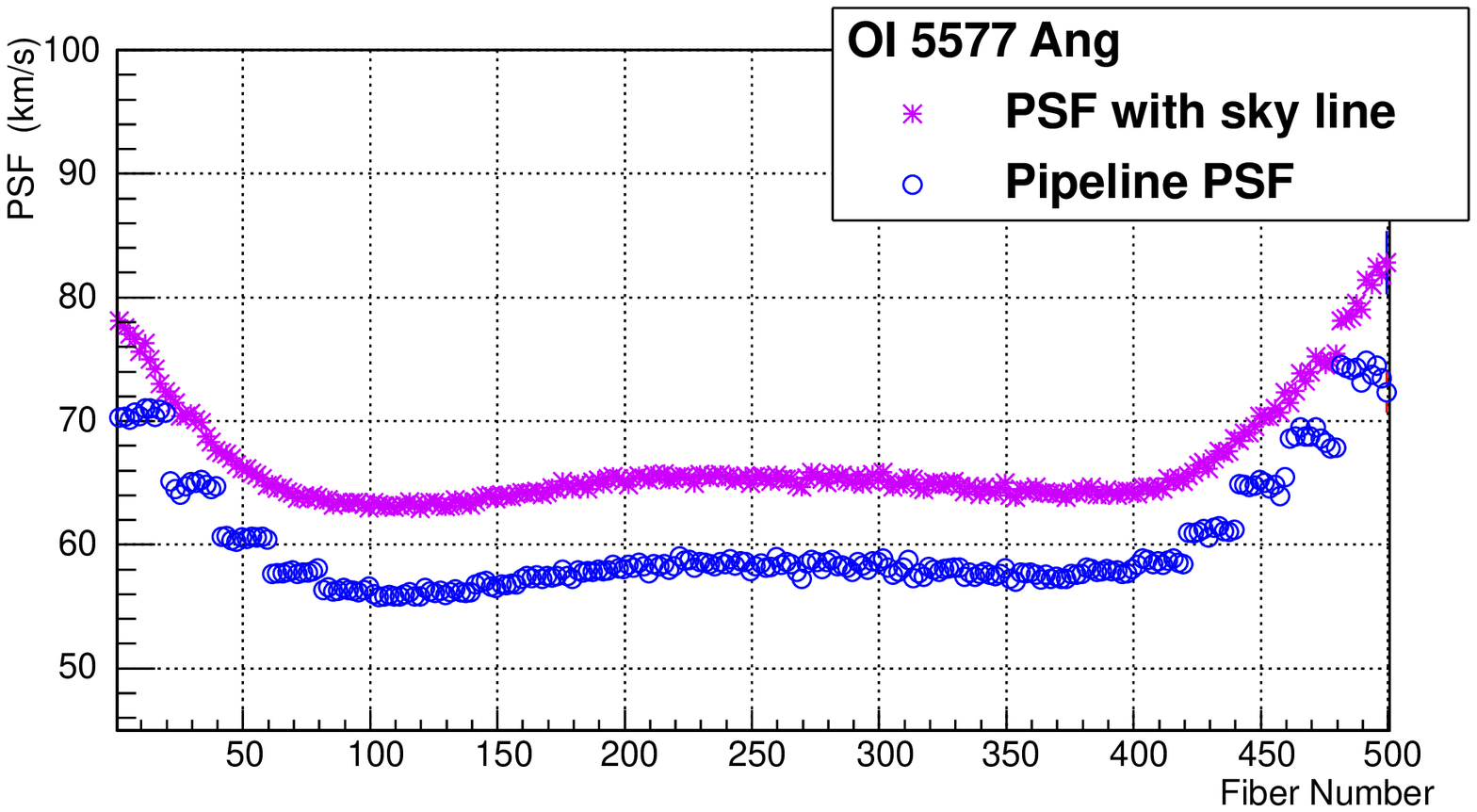, width = \columnwidth} 

\caption[]{Comparison of the resolution given by the pipeline (blue circles) and our computation (purple crosses) for arc lamp or sky line as a function of fiber number (ie. position of spectrum on CCD). Upper left: comparison with arc lamp for a mercury line at $\sim 3650\,\angstrom$. Upper right: comparison with arc lamp for a cadmium line at $\sim 4800\,\angstrom$.   Lower left:  comparison with arc lamp for a mercury line at $\sim 5461\, \angstrom$.  Lower right:  comparison with the OI sky line at $\sim 5577\,\angstrom$. } 
\label{fig:reso}
\end{center}
\end{figure*}

The BOSS reduction pipeline provides the spectrograph resolution  $\sigma_{\lambda,i}$ for each pixel $i$ of each spectrum. On a set of plates, we  performed our own Gaussian fits on the mercury and cadmium lines, and we compared our measurement to the resolution given by the BOSS pipeline. We  observe systematic shifts that depend on two parameters: wavelength (given by the emission wavelength of the line), and  position of the spectrum on the CCD (given by the fiber number). 
Each CCD has 500 fibers, with numbers 1 and 500 corresponding to CCD edges while numbers near 250 correspond to the central region of the CCD.
This comparison is illustrated as a function of fiber number with the first three plots of Fig.~\ref{fig:reso}, corresponding to three  lines of mercury and cadmium. The disagreement  is at most of  a few percent. It is greater in the central region of the CCD and less on the edges. The disagreement increases with  wavelength and reaches 10\%  on the blue CCD, near $\lambda=6000$~\AA.

We  also checked the wavelength calibration using the brightest sky line observed on the blue CCD: the OI line at $\sim 5577\,\angstrom$. The comparison between the BOSS pipeline and our computation of the resolution (see fourth plot of Fig.~\ref{fig:reso}) shows a similar discrepancy as that observed directly with the mercury arc lamp for similar wavelengths.

\subsubsection{Correction of pipeline resolution}

\begin{figure}[htbp]
\begin{center}
\epsfig{figure= 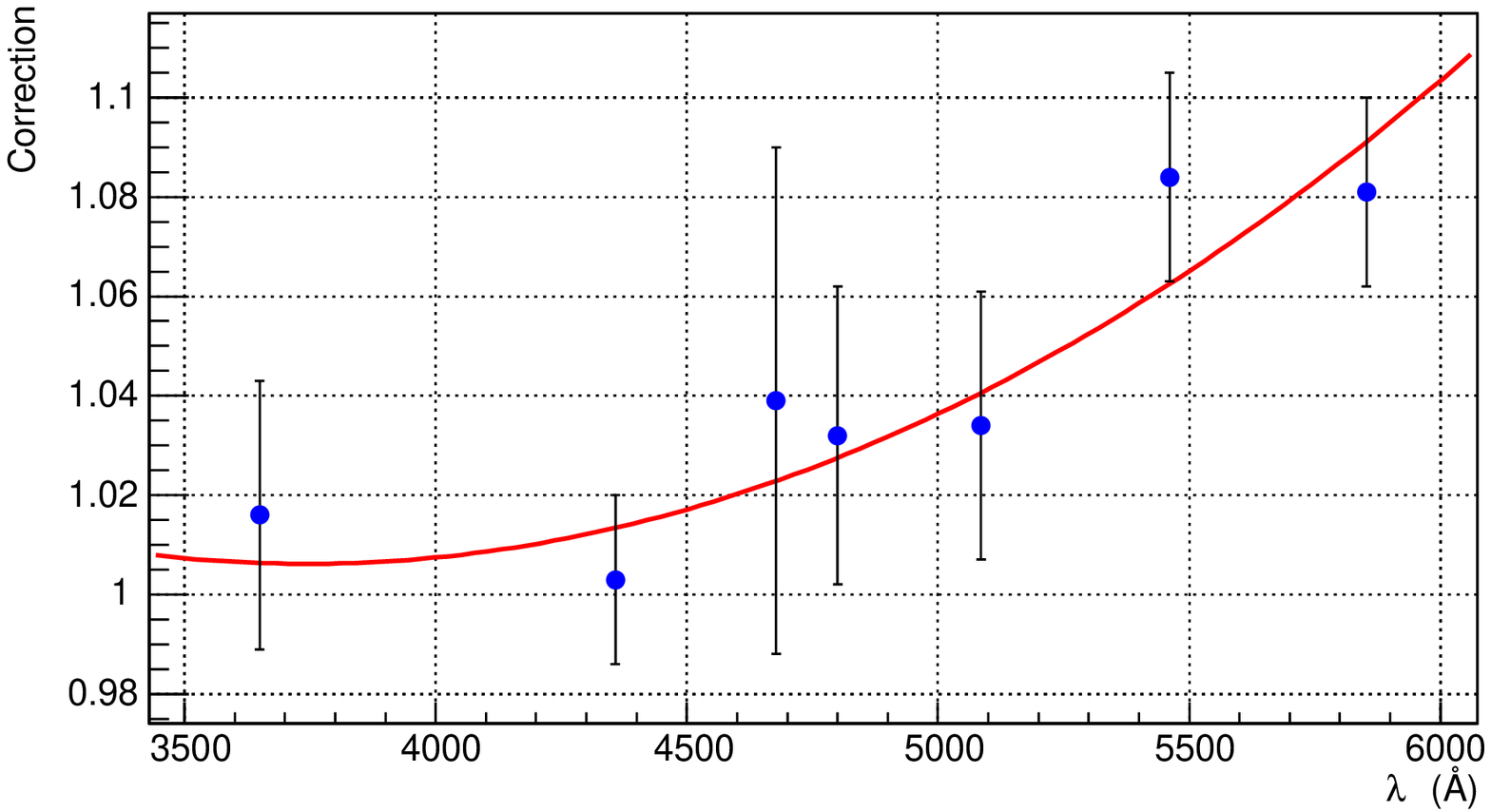,width = \columnwidth} 
\epsfig{figure= 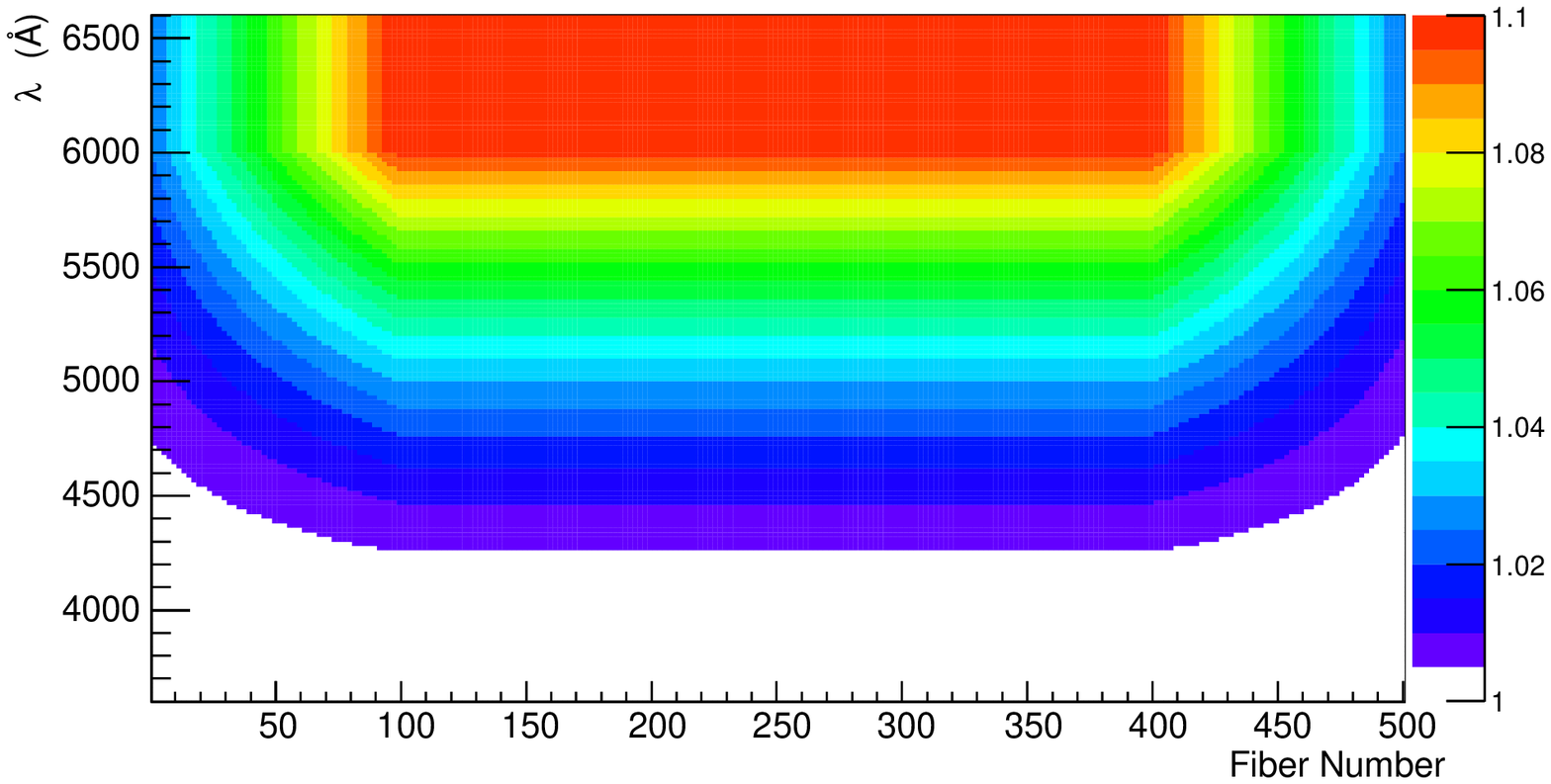,width = \columnwidth} 
\caption[]{\it  
Top: correction of the pipeline resolution for spectra in the middle of the CCD (fiber numbers $\sim 250$). The curve is the best second-order polynomial fit to the measurements at the arc-lamp wavelengths. Bottom: 2D correction table of the pipeline resolution  as a function of  fiber number (ie. position of spectrum on CCD) and  wavelength.  }
\label{fig:resoCor}
\end{center}
\end{figure}

In our analysis, we start from the resolution given by the BOSS reduction pipeline, to which we apply a correction to take   the discrepancy  into account that we  observe   between the pipeline resolution and our estimate, whether with the arc lamp or  a skyline. The top plot of Fig.~\ref{fig:resoCor}  shows the correction as a function of  wavelength for  spectra in the central region of the CCD.  
The amplitude of this correction  is small, on the order of 10\% in the worst case (central spectra and large wavelength for the blue CCD). The bottom plot of Fig.~\ref{fig:resoCor} shows the 2D correction to the resolution that we apply in our analysis, as a function of  wavelength (second-order polynomial) and  fiber number (bounded first-order polynomial).

%%%%%%

\section{Quasar selection and data preparation }\label{sec:data}

\subsection{Data selection}\label{sec:datasel}

We define the Ly$\alpha$ forest by the range $1050<\lambda_{\rm RF}<1180\,$\AA, thus at least 7000 ${\rm km/s}$ away from the quasar Ly$\beta$ and Ly$\alpha$ emission peaks. We limit the spectra to wavelengths above the detector cutoff, i.e., to $\lambda>3650\,$\AA, corresponding to an absorber redshift of $z=2.0$. 

The Ly$\alpha$ forest spans a  redshift range $\Delta z \sim 0.4$ for a quasar at a redshift $z_{\rm qso}= 2.5$, and $\Delta z \sim 0.6$ for a quasar redshift $z_{\rm qso}=5.0$.  To improve our redshift resolution to $\Delta z<0.2$ without  overly affecting  the $k$-resolution and at the same time, to reduce the computation time for the likelihood approach (details in the analysis part, Sec.~\ref{sec:methods}), we split the  Ly$\alpha$ forest into two or three (depending on the length of the Ly$\alpha$ forest)  consecutive and non-overlapping subregions of  equal  length, hereafter called `$z$-sectors'.   
A non-truncated Ly$\alpha$ forest   contains 507 pixels and is divided into three $z$-sectors of 169 pixels each. At low redshift  ($z_{\rm qso}<2.5$) the forest extension is limited by the CCD UV cutoff. In practice, the forest is divided into three $z$-sectors down to a forest length of 180 pixels, into two $z$-sectors for a forest length  between 90 and 180 pixels and not subdivided otherwise. This procedure ensures that the redshift range spanned by a $z$-sector  is  at most 0.2. 

With a pixel size $\Delta v = c\Delta \lambda/\lambda = 69\;{\rm km/s}$, the smallest $k$-mode is therefore between $k_{\rm min}=5\times10^{-4}\;{\rm (km/s)^{-1}}$ and $k_{\rm min}=10^{-3}\;{\rm (km/s)^{-1}}$ depending on the actual $z$-sector length. Our largest possible mode is determined by the Nyquist-Shannon limit at $k_{\rm Nyquist} = \pi/\Delta v = 4.5\times10^{-2}\;{\rm (km/s)^{-1}}$, but we limit our analysis to $k_{\rm max} = 0.02 \;{\rm (km/s)^{-1}}$ because of the large window function correction (mostly due to the spectrograph resolution, cf. Fig.~\ref{fig:window}) for  modes of larger $k$. 

We used the quasars from the DR9 quasar catalog of BOSS~\citep{bib:paris12}. The full catalog contains 61~931 quasars, of which we selected the best 13~821 on the basis of their mean S/N in the Ly$\alpha$ forest, spectrograph resolution ($\overline{R} $), and quality flags on the pixels. 
Flags were also set during the visual scanning of the spectra. We rejected all quasars that have broad absorption line features (BAL),   damped Lyman alpha (DLA)  or detectable Lyman limit systems (LLS)  in their forest. 

The total noise per pixel decreases on average with wavelength by about a factor of 2 between $3650$ and 4000$\,$\AA\ and by another factor of 2 between 4000 and $6000\,$\AA. We reject quasars with S/N$<2$, where the S/N is averaged over the Ly$\alpha$ forest. This criterion mostly removes  low-redshift quasars, since they have their Ly$\alpha$ forest in the blue, hence noisiest, part of the spectrograph. 
The spectrograph resolution $R$ varies slightly with wavelength, from typically $\sim82\,{\rm km/s}$ (at $1\sigma$) at 3650~\AA\ to $\sim 61\,{\rm km/s}$ at 6000~\AA.  It also varies with the position of the spectrum on the CCD (cf. Fig.~\ref{fig:reso}), with a resolution in the central part that is about 7~km/s lower than in the outer regions. 
We reject quasars with a resolution, averaged over the Ly$\alpha$ forest, $\overline{R}>85\;\rm km/s$ to limit the effect of the velocity resolution in the derived power spectrum. We also remove quasars with  pixels in their Ly$\alpha$  forest that are masked by the pipeline   ($<2\%$ of the sample).
The purpose of these restrictions is to ensure that the systematic uncertainty coming from the precision with which the spectrograph noise and resolution can be calibrated remains less than the statistical uncertainty of the estimated power spectra.
These uncertainties will be explained in Sec.~\ref{sec:instrument_syst}. Because both the noise and the resolution are worse in the blue part of the spectrograph, these cuts  affect the low-redshift more than the high-redshift quasars. Since the former are also  much more numerous, we can thus improve the quality of our sample in a region where the systematic uncertainties would otherwise dominate  the statistical ones. 

\begin{figure}[h]
\begin{center}
\epsfig{figure= 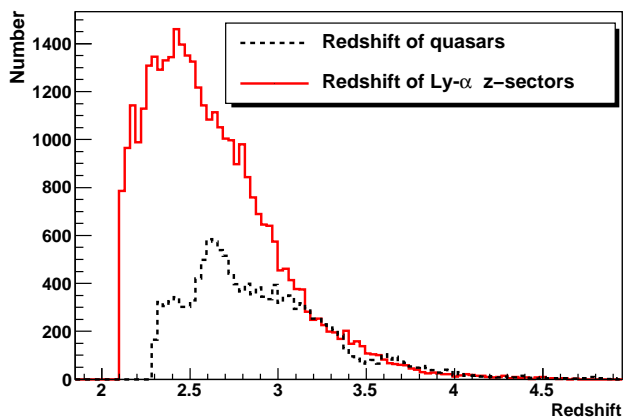,width = \columnwidth} 
\caption[]{\it  Redshift distribution of the $13~821$ quasars selected in the analysis, and mean redshift distribution of each $z$-sector of their Ly${\alpha}$ forest.}
\label{fig:z_QSO}
\end{center}
\end{figure}

Table~\ref{tab:qsocuts} summarizes the impact of our cuts on the quasar sample. Figure~\ref{fig:z_QSO} shows the distributions of the quasar redshifts and of the $z$-sector mean redshifts for the quasars and $z$-sectors that pass these criteria.
\begin{table}[htdp]
\caption{\em Summary of  main quasar selection cuts and fraction of   quasars passing  previous cuts rejected at each step.}
\begin{center}
\begin{tabular}{lc}
\hline\hline
Criteria & Incremental rejection \\
\hline
Mean forest redshift $>$ 2.15 & 46\% \\
S/N$>2.0$ & 36\% \\
$\overline{R}<85\;\rm km/s$ & 40\% \\
Not BAL & 12\%\\
Not DLA & 19\% 
\end{tabular}
\end{center}
\label{tab:qsocuts}
\end{table}

\begin{figure}[h]
\begin{center}
\epsfig{figure= 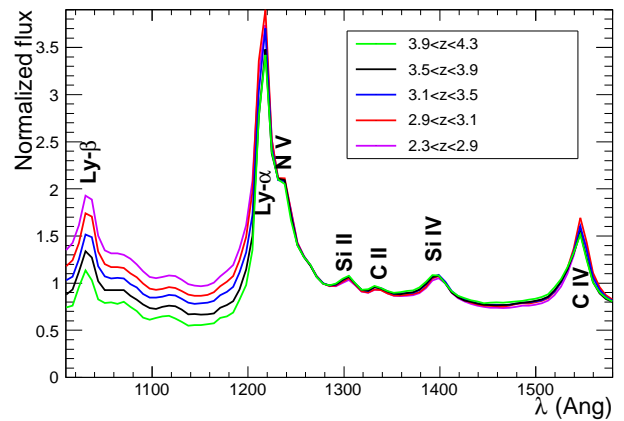,width = \columnwidth} 
\caption[]{\it  Average quasar spectra in five redshift bins. All spectra are normalized at $\lambda = 1280$\AA.}
\label{fig:qso_flux}
\end{center}
\end{figure}

In Fig.~\ref{fig:qso_flux}, we show   average quasar spectra obtained by averaging the spectra of all the DR9 BOSS quasars passing the above cuts, split into five redshift bins from $z=2.3$ to 4.3. Broad quasar emission lines are clearly visible, such as Ly$\beta$ at $\lambda_{\rm RF}\sim 1026$\AA, Ly$\alpha$ at $\lambda_{\rm RF}\sim 1216$\AA, \ion{N}{v}  at $\lambda_{\rm RF}\sim 1240$\AA, \ion{Si}{iv}  at $\lambda_{\rm RF}\sim 1400$\AA\, and \ion{C}{iv}  at $\lambda_{\rm RF}\sim 1549$\AA. The absorption by Ly$\alpha$ absorbers along the quasar line of sight appears blueward of the quasar Ly$\alpha$ emission peak, with more absorption (and thus less transmitted flux) at high redshift. 

We calculate the 1D power spectra in twelve redshift bins of width $\Delta z$=0.2 and centered on $z_c=2.2$ to $z_c=4.4$. The mean redshift of the Ly$\alpha$ absorbers of a given $z$-sector determines the redshift bin to which it contributes. While the Ly$\alpha$ forest of a quasar spectrum may cover several redshift bins, a given $z$-sector only contributes to a single bin, thus avoiding correlations between redshift bins. The redshift span of a $z$-sector, at most 0.2, is  adapted well to the size of our redshift bins.

\subsection{Sky line masking}\label{sec:skylinemask}
Sky lines  affect the data quality by  increasing significantly the pixel noise. The procedure used to identify them is detailed in~\citet{bib:KG13}. We briefly summarize it  here. 

We use the sky calibration fibers and compute the mean and the rms of the  residuals measured on the sky-subtracted spectrum  obtained with the standard BOSS pipeline. We define a `sky continuum' as the running average of the residual rms fluctuation centered on a $\pm 25$ pixel window, and  generate a list of sky lines from all the wavelengths that are above 1.25~$\times$ the sky continuum. The continuum, measured with the unmasked pixels, and the sky line list are iterated until they converge. To this list, we add the calcium H and K  Galactic absorption lines near $\lambda = 3933.7\,$\AA\ and $\lambda = 3968.5\,$\AA. We then mask all pixels that are within 1.5~\AA\ of the listed wavelengths.

We apply the mask differently  in the FT and the likelihood methods. For the FT,  we replace the flux of each masked pixel by the average value of the flux over the unmasked forest. This procedure introduces a $k$-dependent bias in the resulting power spectrum that  reaches at most 15\% at small $k$ for the $3.5<z<3.7$ redshift bin,  which contains 5577~\AA\ OI, the strongest sky emission line. We correct for this bias a posteriori, as explained in  Sec.~\ref{sec:methods_syst}. For the likelihood method, the masked pixels are simply omitted from the data vector. We have checked (see details in Sec.~\ref{sec:methods_syst}) that in this case  we observe no bias on the resulting power spectrum.

\subsection{Quasar continuum}\label{sec:cont}
The normalized transmitted flux  fraction $\delta(\lambda)$ is estimated from the pixel flux $f(\lambda)$ by:
\begin{equation}
\delta(\lambda) = \frac{f(\lambda)}{f_{\rm qso}^{1280}  C_q(\lambda,z_{\rm qso}) \bar{F}(z_{\rm Ly\alpha})} - 1\, ,
\label{eq:delta}
\end{equation}
where $f_{\rm qso}^{1280}$ is a normalization  equal to the mean flux in a $20\,$\AA\ window about $\lambda_{RF} = 1280\,$\AA, $C_q(\lambda,z_{\rm qso})$ is the normalized unabsorbed flux (the mean quasar `continuum') and $\bar{F}(z_{\rm Ly\alpha})$ is the mean transmitted flux fraction at the \ion{H}{i} absorber redshift. Pixels affected by sky line emission are not included when computing the normalization. Since the mean quasar continuum is flat in the normalization region, the rejection of a few pixels does not bias the mean pixel value. The product $ C_q(\lambda,z_{\rm qso}) \bar{F}(z_{\rm Ly\alpha})$ is assumed to be universal for all quasars at redshift $z_{\rm qso}$ and is computed by stacking appropriately normalized quasar spectra $f/f_{\rm qso}^{1280}$, thus averaging out the fluctuating Ly$\alpha$ absorption.  The product $f_{\rm qso}^{1280}  C_q(\lambda,z_{\rm qso}) \bar{F}(z_{\rm Ly\alpha})$ represents the mean expected flux, and   the transmitted flux fraction is given by $F=f / (f_{\rm qso}^{1280}C_q)$. For a pixel at rest-frame wavelength $\lambda_{\rm RF}$ of a quasar at redshift $z_{\rm qso}$, the corresponding \ion{H}{i} absorber redshift $z_{\rm Ly\alpha}$ can be inferred from $1+z_{\rm Ly\alpha} = \lambda_{\rm RF} / \lambda_{\rm Ly\alpha}\times(1+z_{\rm qso})$, where $\lambda_{\rm Ly\alpha} \simeq 1216$\AA.

\begin{figure}[h]
\begin{center}
\epsfig{figure= 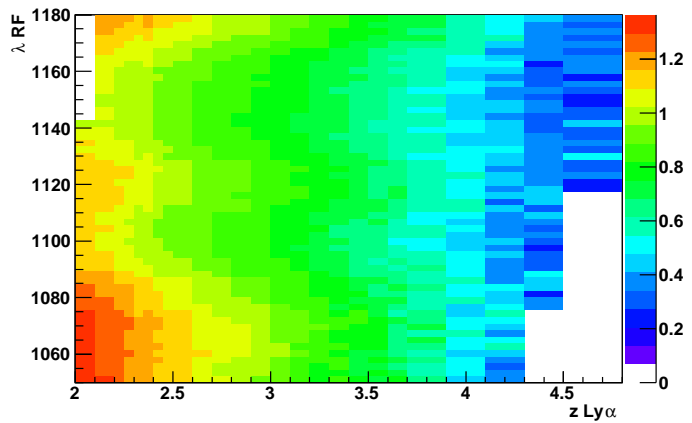,width = \columnwidth} 
\caption[]{\it Product  of the quasar continuum $C_q(\lambda, z_{\rm qso})$ by the mean transmitted flux fraction $\bar{F}(z)$ as a function of  rest-frame wavelength and Ly$\alpha$ redshift. This 2D table is used to compute the normalized flux transmission fraction $\delta(\lambda)$.}
\label{fig:continuum2D}
\end{center}
\end{figure}
\begin{figure}[h]
\begin{center}
\epsfig{figure= 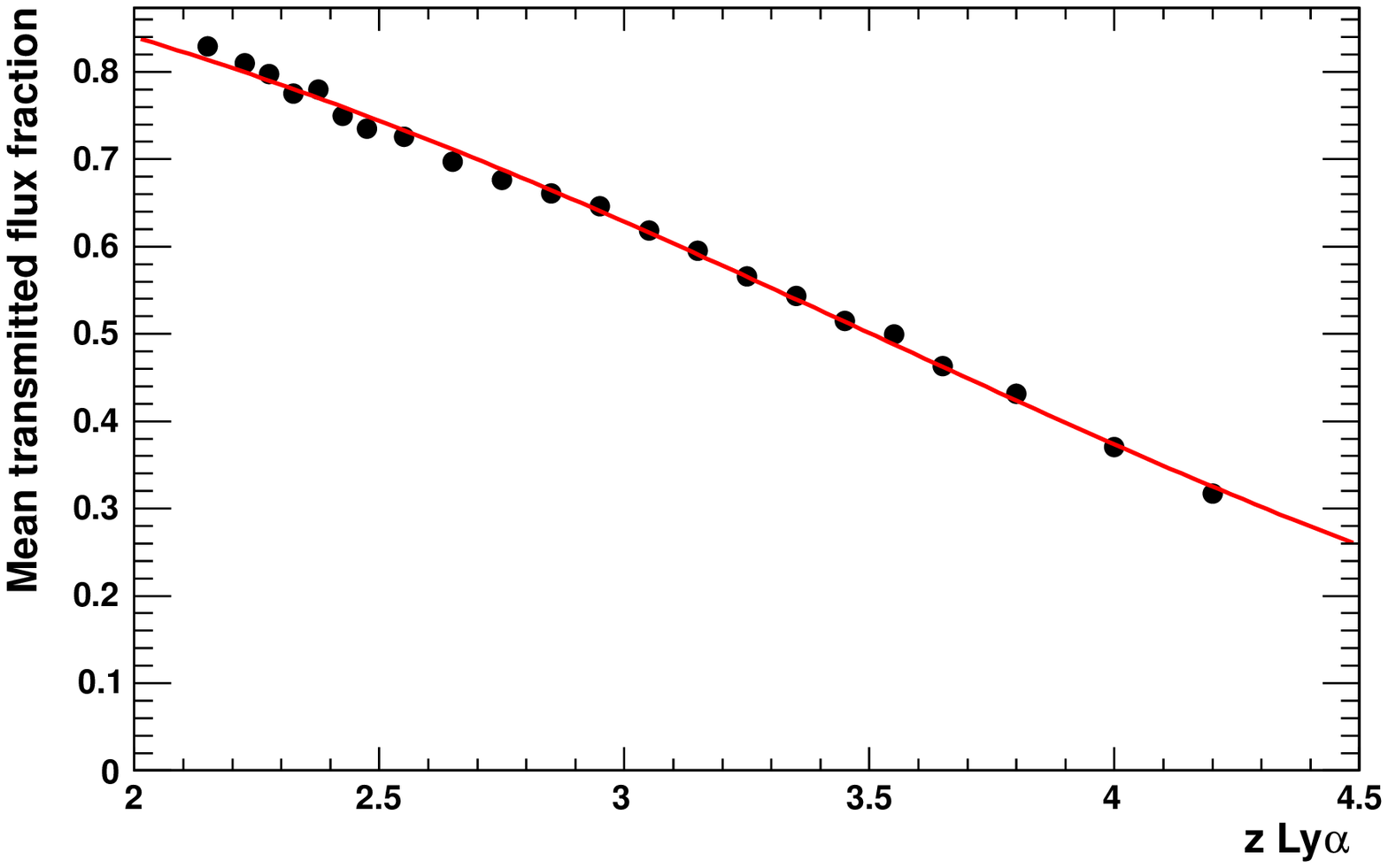,width = \columnwidth} 
%\caption[]{\it  Mean transmitted fraction $\bar{F}(z)$ as a function of   Ly$\alpha$ redshift, relative to the quasar flux at $\lambda_{RF} = 1280\;\AA$. The overlaid curve is $1.35\times \exp [-0.0043(1+z)^{3.3}]$.}
%\label{fig:LUT_abso}
\epsfig{figure= 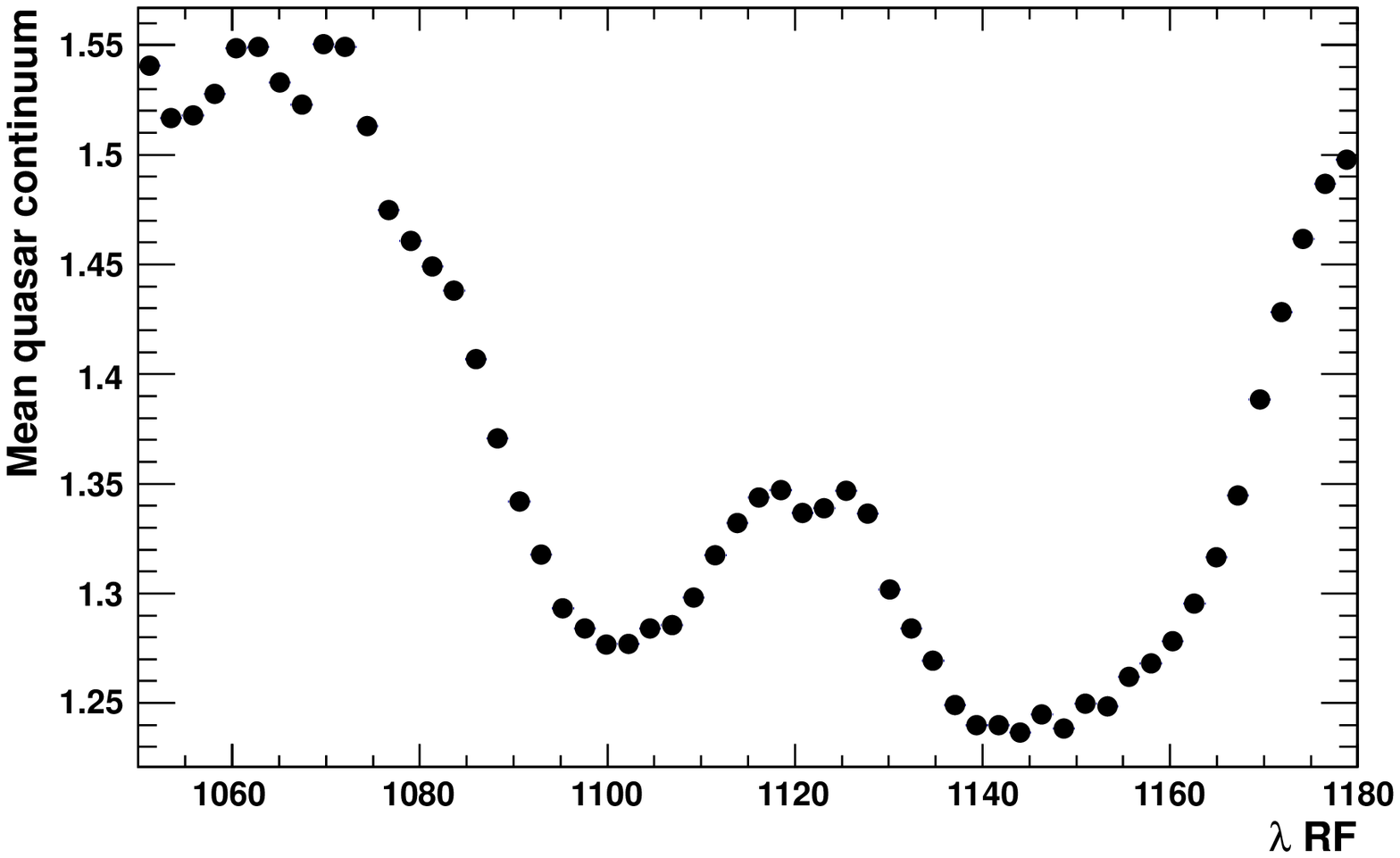,width = \columnwidth} 
\caption[]{\it Top: mean transmitted fraction $\bar{F}(z_{\rm Ly\alpha})$ as a function of   Ly$\alpha$ redshift. The overlaid curve is $\exp [-0.0046(1+z)^{3.3}]$. Bottom: mean quasar continuum $C_q(\lambda)$ as a function of  rest-frame wavelength, averaged over all selected  quasars.}
\label{fig:LUT_proj}
\end{center}
\end{figure}

Figure~\ref{fig:continuum2D} shows the product $C_q(\lambda,z_{\rm qso}) \bar{F}(z_{\rm Ly\alpha})$ of the quasar continuum with the mean transmitted flux fraction as a function of  rest-frame wavelength and Ly$\alpha$ redshift. Figure~\ref{fig:LUT_proj} shows the projection of the 2D distribution of Fig.~\ref{fig:continuum2D} onto the redshift  or the  wavelength axis. The former shows $\langle f(\lambda)/f_{\rm qso}^{1280}) \rangle$ averaged over wavelength and is proportional to  the mean transmitted flux fraction, and the latter shows the mean unabsorbed quasar spectrum $C_q(\lambda)$ normalized to $f_{\rm qso}^{1280}$. The mean transmitted flux fraction is well fit by a function of the form $\exp [-\alpha(1+z)^{\beta}]$, with $\alpha \sim 0.0046$ and $\beta \sim 3.3$, in agreement with previous measurements of the optical depth $\tau_{\rm eff}$ where $\bar{F} \propto \exp(-\tau_{\rm eff})$ (see e.g. \citep{bib:meiksin09} for a review).

The values in the 2D table, $C_q(\lambda,z_{\rm qso}) \bar{F}(z_{\rm Ly\alpha})$, differ from those of the product $C_q(\lambda) \bar{F}(z_{\rm Ly\alpha})$ by up to 5\%, possibly due to variations in the mean quasar continuum with redshift. Despite its lower statistical precision for a given wavelength and redshift, we therefore  use the 2D table.
%Because we note up to 5\% differences  between the values in the 2D table and the product $C_q(\lambda) \bar{F}(z)$, possibly due to variations in the mean quasar continuum with redshift, we  use the 2D table, i.e.  $C_q(\lambda,z_{\rm qso}) \bar{F}(z_{\rm Ly\alpha})$. 

\begin{figure}[h]
\begin{center}
 \epsfig{figure= 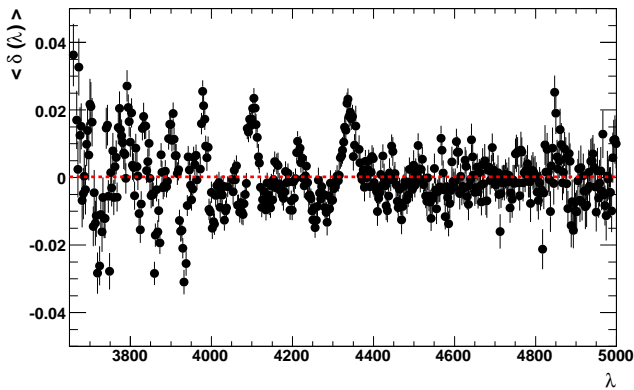,width = \columnwidth} 
\caption[]{\it Mean of $\delta(\lambda)$ as a function of wavelength in \AA. Systematic offsets from zero are seen at the 2\% level due to imperfections in the spectrograph calibration.}
\label{fig:balmer}
\end{center}
\end{figure}

Figure~\ref{fig:balmer} shows the resulting mean $\delta$ as a function of observed wavelength. The mean fluctuates about zero at the 2\% level with correlated features that are due to imperfect spectrograph calibration and absorption.  These features include the calcium H-K doublet at (3934, 3968\AA) from Milky Way absorption, and Balmer lines H$\gamma$, $\delta$, $\epsilon$ at (4341, 4102, 3970 \AA) that are residuals from the use of F-stars as spectrocalibration standards. \citet{bib:busca13} have studied these features in detail and concluded that they had quasar-to-quasar variations of less than 20\%  of the mean Balmer artifact deviations.
To remove their contribution to the Ly$\alpha$ power spectrum, we subtract   the mean residual of Fig.~\ref{fig:balmer} from $\delta(\lambda)$. 
 
%%%%%%

\section{Methods for determining $P(k)$}\label{sec:methods}
We apply two  methods to compute the one-dimensional power spectrum. The first  one  is  based on a FT. It is fast and robust, thus
allowing many tests leading to a better understanding of the impact of
the different ingredients entering the analysis. We use it to test the impact of, for instance, different selections of quasars on the precision of the resulting power spectra or  various algorithms to mask sky emission lines. The second method relies upon  a maximum likelihood estimator in real space. It can take  variations in the noise or in the spectrograph resolution at the pixel level into account instead of through global factors, and is therefore expected to be more precise than a FT. It also offers a natural way to mask pixels affected by sky emission lines, as explained in Sec.~\ref{sec:method_lkl}. However, it  is more sensitive than the FT to details in the implementation of the method, is susceptible to convergence problems in the presence of noisy
spectra and is more time-consuming. It is therefore not as flexible for algorithm testing. The 
power spectra obtained with the two approaches are in good agreement. Their  comparison provides an estimate of the systematic uncertainty on our measurement (cf. Sec.~\ref{sec:resultsummary}). 

\subsection{Fourier transform approach}
\label{sec:fft}

\subsubsection{Measurement of the power spectrum with a Fourier transform}

To measure the 1D power spectrum $P_{1D}(k)$  we decompose each absorption spectrum $\delta_{\Delta v}$ into Fourier modes  and  estimate their variance as a function of wave number.  In practice, we do this  by computing the discrete FT of  the  flux transmission fraction $\delta=F/\langle F\rangle - 1$  as described in~\citet{bib:croft98}, using a fast Fourier transform  (FFT) algorithm. Using a FFT requires that the pixels be equally spaced. This condition is satisfied with the quasar coadded spectra provided by the SDSS pipeline~\citep{bib:bolton12}: the spectra are computed with a constant pixel  width  $\Delta[\log(\lambda)] = 10^{-4}$, and the velocity difference between pixels, i.e., the relative velocity of absorption systems at wavelengths $\lambda+\Delta\lambda/2$ and $\lambda-\Delta\lambda/2$, is $\Delta v =  c\, \Delta \lambda / \lambda = c\, \Delta [\ln(\lambda)]$. The coadded spectra thus have  equally spaced pixels in $\Delta v$. 
Throughout this paper we  therefore use velocity instead of observed wavelength. Similarly, the wave vector $k\equiv 2\pi/ \Delta v$ is measured in $\rm(km/s)^{-1}$.

In the absence of instrumental effects (noise and resolution of the spectrograph), the 1D power spectrum can be simply written as the ensemble average over quasar spectra of $P^{raw}(k) \equiv   \left| \mathcal{F}(\delta_{\Delta v})  \right|^2$, where $\mathcal{F}(\delta_{\Delta v})$ is the FT of the normalized flux transmission fraction $\delta_{\Delta v}$  in the quasar Ly$\alpha$ forest binned in pixels of width $\Delta v$.

\begin{figure}[h]
\begin{center}
\epsfig{figure= 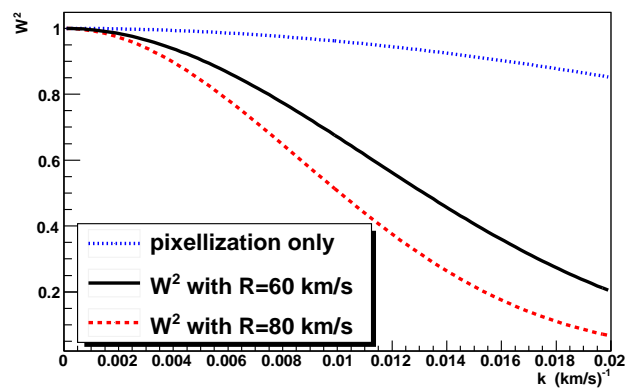,width = \columnwidth} 
\caption[]{\it  Window function $W^2(k,\bar{R},\Delta v)$, with $\Delta v=69\,{\rm km/s}$, reproducing the spectrum binning and the impact of the spectrograph resolution, for a  resolution $\bar{R}=60\,{\rm km/s}$ typical at $\lambda>5000\,$\AA\ and  $\bar{R}=80\,{\rm km/s}$ typical at $\lambda<4300\,$\AA. For comparison, we also show the contribution from only the pixellization   (equivalent to $\bar{R}=0$).}
\label{fig:window}
\end{center}
\end{figure}

When taking  the noise in the data and the impact of the spectral resolution of the spectrograph into account, $\delta$ can be expressed as $\delta= s + n$, with $s$ the  signal and $n$ the noise, and the estimator of the 1D power spectrum is
\begin{equation}
P_{1D}(k)  =  \left< \frac{ P^{raw}(k) - P^{noise}(k) }{W^2(k,R,\Delta v)} \right> \, ,
\label{eq:P1D_FFT}
\end{equation}
where  $\langle\rangle$ denotes the ensemble average over quasar spectra and where
\begin{equation}
P^{noise}(k) \equiv  \left| \mathcal{F}(n_{\Delta v}) \right|^2 \; .
\end{equation}
The window function corresponding to the spectral response of the spectrograph is defined by
\begin{equation}
W(k,R,\Delta v)= \exp\left(- \frac{1}{2}(kR)^2\right) 
	\times \frac{\sin(k\Delta v /2)}{(k\Delta v/2)}\, ,
\label{eq:window}
\end{equation}
where $\Delta v$ and $R$ are  the  pixel width and the spectrograph resolution, respectively. Both quantities are in $\rm km/s$, and $R$ should not  be confused with the dimensionless resolving power of the spectrograph. We illustrate in Fig.~\ref{fig:window} the  spectrograph resolution on the window function $W^2(k,\bar{R},\Delta v)$ for different values of $\bar{R}$. 

\subsubsection{Computation of  $P_{1D}(k)$ with a FFT}

We compute the FT using the  efficient FFTW package\footnote{http://www.fftw.org, 
The FFTW package was developed by M.~Frigo and S.~G.~Johnson, 1998.}. Compared to the likelihood approach described in the next section, the Fourier transform is much faster, but it requires some simplifying hypotheses in the treatment of the noise and of the spectrograph resolution. We explain these simplifications below.  Sky emission lines are also treated in a simplified way as described in Sec.~\ref{sec:skylinemask}. The results provided by this simple method are  complementary to the likelihood approach.

Although the redshift of the absorbing hydrogen  increases with wavelength along the spectrum of a given quasar, the power spectrum is considered to be computed at their average redshift. As explained in Sec.~\ref{sec:datasel}, we improve the redshift resolution of the measured power spectra by splitting the Ly$\alpha$ forest of each quasar into redshift subregions (or $z$-sectors, see Sec.~\ref{sec:datasel}). The computation is done separately on each $z$-sector instead of  on the entire Ly$\alpha$ forest. The mean redshift of the Ly$\alpha$ absorbers in the $z$-sector determines the redshift bin to which the $z$-sector contributes.

The noise power spectrum $P^{noise}(k,z) $ is taken as the  power spectrum $P_{\rm diff}^{noise}$  on the $z$-sector, computed as explained in Sec.~\ref{sec:noise}. Since $P_{\rm diff}^{noise}$  is flat with $k$, we improve the statistical precision on our determination of the level of the noise power spectrum by taking the average of  $P_{\rm diff}^{noise}(k)$ for $k<0.02\,{\rm (km/s)}^{-1}$. 

Finally, we apply the correction of the spectrograph resolution by dividing by $W^2(k,\overline{R},\Delta v)$, where $\overline{R}$ is the mean value of the spectral resolution $R$ averaged over the $z$-sector. The value of $R$ is given by the pipeline and corrected following the prescription described in Sec.~\ref{sec:psf}.
For a given spectrum,  $R$ varies by less than 10\% over the  Ly$\alpha$ forest (less than 3\% over a $z$-sector), and the impact of this simplification is  negligible.  

We rebin the final power spectrum onto a predefined grid in $k$-space, giving equal weight to the different Fourier modes that enter each bin. The final 1D power spectrum is obtained by averaging the corrected power spectra of all the contributing $z$-sectors of all selected quasars, as expressed in equation~\ref{eq:P1D_FFT}.

\subsection{Likelihood approach}\label{sec:method_lkl}
\label{sec:lkl}

We estimate $P_{1D}(k)$ using a maximum likelihood estimator derived from methods developed for studies of the cosmic microwave background anisotropy~\citep{bib:bond98, bib:seljak98}. This method guarantees optimal performance for Gaussian or nearly Gaussian distributions, and can  be applied here ensuring minimal variance, although the power spectrum estimates are not Gaussian distributed. Our approach involves a direct maximization of the likelihood function and is not based on the quadratic maximum estimation as in~\citet{bib:mcdonald06}.  It is slower but provides the  values of $P_{1D}(k)$  with their covariance matrix at the maximum of the  likelihood.

\subsubsection{The likelihood function}

We model the  normalized flux transmission fraction $\delta_i = F_i/ \langle F \rangle - 1$ measured in pixel $i$ as contributions from signal and noise: $\delta_i = s_i + n_i$. We assume that signal and  noise are independent, with zero mean and covariance matrices given by
\begin{equation}
C_{ ij}^S=\langle s_i s_j \rangle \;\;{\rm and} \;\; C_{ ij}^N=\langle n_i n_j \rangle=\sigma_i\sigma_j \delta_{ij} \;,
\end{equation}
where $ \delta_{ij}$ is the Kronecker symbol and $\sigma_i = \sigma_p/\alpha_{\rm cor}^{noise}$ (pipeline estimate and its correction).
The total covariance matrix can therefore be written as
\begin{equation}
C=\langle \delta_i \delta_j \rangle = C_{ ij}^S + C_{ ij}^N \; .
\end{equation}
The signal covariance matrix can be derived from the 1D power spectrum by 
 \begin{eqnarray*}
C_{ ij}^S  & = & \int_{-\infty}^{+\infty}P_{1D}(k)\cdot\exp\left[ -ik\Delta v \times (i-j)\right]  dk \\ 
	& = &\int_{0}^{+\infty}P_{1D}(k)\cdot 2\cos\left[ k\Delta v \times (i-j) \right] \, dk . \\
\end{eqnarray*}
We can approximate $P_{1D}$ by $\mathbf{P} = (P_1,... P_\ell ..., P_{N_\ell})$, a discrete set of  $N_\ell$ values of $P_\ell\equiv P_{1D}\left(\frac{k_\ell+k_{\ell-1}}{2}\right)$ for the modes $k_\ell$. The previous integral can then be approximated by
\begin{eqnarray}
C_{ ij}^S(\mathbf{P} )  & = & \sum_{\ell=1}^{N_\ell} P_\ell 
	\cdot \int_{k_{\ell-1}}^{k_\ell} 2\cos\left[ k\Delta v\times (i-j)\right] \, dk . 
\end{eqnarray}

Taking the spectrograph resolution into account and using the definition of the window function given in Eq.~\ref{eq:window}, the covariance matrix becomes
\begin{eqnarray*}
C_{ ij}^S(\mathbf{P} )  =  \sum_{\ell=1}^{N_\ell} P_\ell 
	\cdot \int_{k_{\ell-1}}^{k_\ell} && 2  \cos\left[ k \Delta v\times (i-j)\right] \times \\
	& & W(k,R_i,\Delta v) W(k,R_j,\Delta v)\, dk 
\end{eqnarray*}
where $R_i$ and $\Delta v$ are respectively the spectrograph resolution for pixel $i$ and the pixel width (same for all pixels).

For  spectrum $sp$ containing $N_{sp}^{\rm pix}$ pixels, we can define the likelihood function $\mathcal{L}_{sp}$  as
\begin{equation}
\mathcal{L}_{sp} (\mathbf{P} ) = \frac{1}{(2\pi)^{N_{sp}^{\rm pix}/2} \sqrt{\det(C)}}\exp \left(-\frac{\delta^T C^{-1} \delta}{2} \right) .
\end{equation}
For stability reasons, we  do not fit a single spectrum at a time but instead  combine  ${N_{sp}}$ spectra corresponding to the same redshift bin into a common likelihood. The likelihood is the product:
\begin{equation}
\mathcal{L}(\mathbf{P} (z)) = \prod_{{sp}=1}^{N_{sp}} \mathcal{L}_{sp}(\mathbf{P} (z)).
\label{eq:likelihood}
\end{equation}
We can then search for the vector $\mathbf{P} (z)$ (i.e. the parameters $P_\ell(z)$) that maximizes this likelihood.  

\subsubsection{Extraction of  $P_{1D}(k)$}

We extract the $P_{1D}(k)$ power spectrum from the likelihood $\mathcal{L}(\mathbf{P}(z))$  that combines several spectra in the same redshift bin. We use the MINUIT~\citep{bib:minuit}  package to minimize the term $-2\ln(\mathcal{L})$. This minimization provides the value and the error of  each $P_\ell(z)$ and the covariance matrix between the different $P_\ell(z)$. The method implemented  in MINUIT package may be slow but it is  robust, because it  seldom falls into  secondary minima. 

As the minimization can take a few hundred iterations, we have optimized our fitting procedure. The computation time of the likelihood is limited by the inversion of the covariance matrix $C$. Therefore, to reduce the size of the matrix $C$ (number of pixels), we do the computation on the `$z$-sectors' defined in Sec~\ref{sec:datasel}, instead of on the entire  Ly$\alpha$ forest. 

The noise  covariance matrix is assumed diagonal, i.e., without correlation terms. Each diagonal element is equal to the  square of the pixel error estimated by the pipeline, $\sigma_p$, multiplied by the square of the correction  factor $\alpha_{\rm cor}^{noise}$ defined in Eq.~\ref{eq:corr}.

We use the Cholesky decomposition to increase the speed of the matrix inversion of the positive-definite matrix $C$. The Cholesky decomposition is roughly twice as efficient as the $LU$ decomposition, and it is numerically more precise. 

Finally, in the product of the individual likelihoods of Eq.~\ref{eq:likelihood}, we take $N_{sp}=100$ where in practice $N_{sp}$ is  the number of  $z$-sectors  and not the number of quasar spectra. While a large $N_{sp}$ improves  the fit convergence by making the fit more stable, we nevertheless restrict the number of $z$-sectors to be fitted simultaneously in order to limit the minimization to a reasonable CPU time. 
We determine the final $\mathbf{P}(z)$ by averaging over  the $N_b$ bunches of $N_{sp}$ $z$-sectors (with $N_b\times N_{sp}$ being the total number of $z$-sectors that enter a given redshift bin). The total covariance matrix $M_{\rm cov}^{tot}$ is  computed as $(M_{\rm cov}^{tot})^{-1}=\sum (M_{\rm cov}^{b})^{-1}$.

The typical CPU time for the minimization of one bunch of 100  $z$-sectors is about 10 to 15 minutes,
performing between 500 and 600 iterations before convergence.  For this analysis, we ran on a farm of
24 computers, which allowed us to compute the independent power spectra for different redshift bins in parallel. The total wall-clock time for the full analysis is approximately 12 hours.

%%%%%%

\section{Systematic uncertainties}\label{sec:systs}
In this section, we study the biases and systematic uncertainties that affect our analysis. We correct our result for the identified biases, and we estimate  systematic uncertainties that we summarize in Sec.~\ref{sec:results} (Tables~\ref{tab:Resultsfft} and \ref{tab:Resultslkl}), along with our measured power spectrum.

The biases and uncertainties arise from two different origins. In Sec. \ref{sec:methods_syst}, the biases related to the analysis methods, either FT or likelihood, are presented assuming that the instrumental noise and resolution are perfectly known.  Then in Sec. \ref{sec:instrument_syst},  the systematics due to our imperfect knowledge of the instrument characteristics are described and quantified using the data themselves. 

\subsection {Biases in the analyses and related systematics}
We study here the  biases and systematic uncertainties introduced at each step of the data analysis. We estimate their impact using  mock spectra. We compute the `bias'  of the method as the ratio of the measured flux power spectrum to the flux power spectrum that was generated in the mock spectra. 

We generated mock spectra with the following procedure. First, a redshift and a $g$-magnitude are chosen at random from the real BOSS spectra. Second, an unabsorbed flux spectrum is drawn for each quasar from a random selection of PCA amplitudes following the procedure of~\citet{bib:paris11} and flux-normalized to  the selected $g$ magnitude. Third, the Ly$\alpha$ forest absorption is generated following a procedure adapted from~\citet{bib:font-ribera12}. They provide an algorithm for generating any spectrum of the transmitted flux fraction $F(\lambda)$ from a Gaussian random field $g(\lambda)$. Specifically, they present a recipe for choosing the parameters $a$ and $b$ and the power spectrum $P_g(k)$ such that the transformation $F(\lambda)=\exp[-a\exp(bg(\lambda))]$ yields the desired power spectrum and mean value of $F(\lambda)$. In practice we generate a suite of transmitted-flux-fraction spectra 
for twelve redshifts that reproduce the observed power. For each wavelength pixel,  $F(\lambda)$ is  obtained by interpolation between redshifts according to the actual Ly$\alpha$ absorption redshift of the pixel. The unabsorbed flux is  multiplied by $F(\lambda)$ and convolved with the spectrograph resolution. In practice,  the spectra are  generated with a pixel width that is one third of an SDSS pixel, and about one third of the spectral resolution. We checked that this size was  small enough to  take  the spectral resolution into account properly. Finally,  noise is added according to BOSS throughput and sky noise measurements as was done in~\citet{bib:legoff11}, and the spectrum is rebinned  to the SDSS bin size.

The determination of the transmitted flux fraction requires an estimate of the quasar unabsorbed flux  obtained as explained in Eq.~\ref{eq:delta} of Sec.~\ref{sec:cont}. As a  starting point, we have checked that using the generated values for the quasar continuum $C_q(\lambda)$ and for the mean transmitted flux $\bar{F}(z)$ allows  recovery of  exactly the input power spectra in the absence of noise. Using instead our estimated value of $C_q(\lambda,z_{\rm qso}) \bar{F}(z)$ produces an overestimate of the power spectrum of  order 2\%, and  is $k$-independent over the $k$-range of interest. To have a better estimate of the continuum on a quasar-by-quasar basis and allow for tilts in the flux calibration, we  considered an improved method  consisting of multiplying the average shape by a factor $A+B\lambda_{\rm RF}$  where $A$ and $B$ were fitted for each quasar. This method was not retained, however, because it generated a larger overestimate ($\sim 6\%$). 

We  studied the impact of our correction for the spectrograph spectral resolution $W(k, R,\Delta v)$ by using mocks where $W$ was either similar to that of BOSS (including both pixellization and spectrograph resolution) or  reduced to the  contribution of  pixellization alone (cf. Fig.~\ref{fig:window}). We found negligible bias (less than 0.1\%) in both the FT and the likelihood methods.

The removal of the noise contribution to the Ly$\alpha$ power spectrum  introduces a bias in both methods. For mock spectra, the noise power spectrum is white, and we determine its level  directly  from the pixel errors.  For the FT approach, the removal of the noise power spectrum on mock spectra analyzed with the true quasar continuum produces a small ($2\%$) underestimate. 

\begin{figure}[h]
\begin{center}
\epsfig{figure= 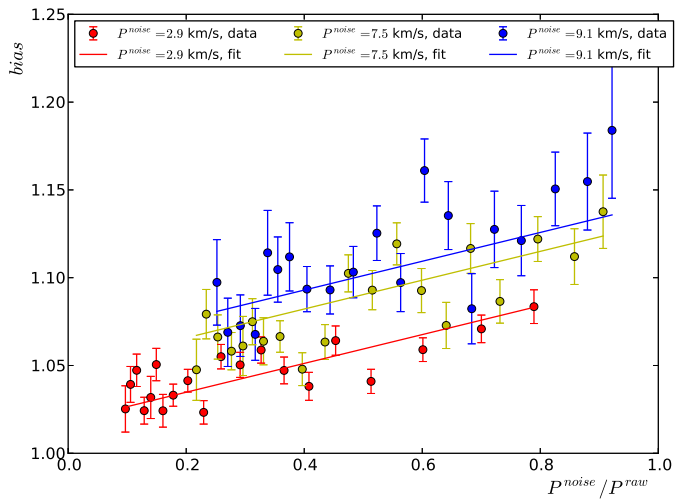,width = \columnwidth} 
\caption[]{\it  Overestimate of the mock power spectrum determined from  the likelihood method  as a function of $P^{noise}/P^{raw}$, for different values of $P^{noise}$. The curves illustrate the best fit model.}
\label{fig:bias_LKL}
\end{center}
\end{figure}

\begin{figure}[h]
\begin{center}
\epsfig{figure= 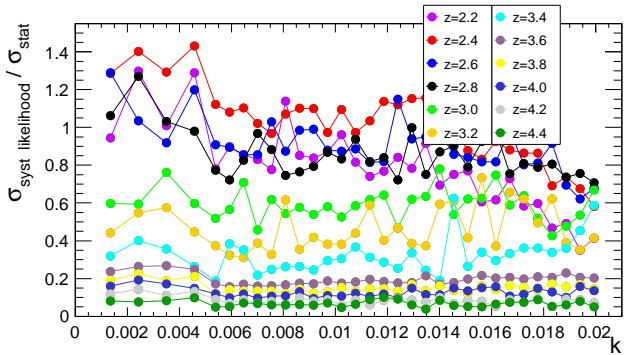,width = \columnwidth} 
\caption[]{\it Systematic uncertainty related to the correction of the noise-related bias in the likelihood method,   relative to the statistical uncertainty, for each redshift bin.}
\label{fig:systLKL}
\end{center}
\end{figure}

The likelihood method is much more sensitive than the Fourier transform approach to the  level of noise and to the relative levels of noise and signal power spectra. It results in biases that can reach $\sim 13\%$ at low redshift ($z<2.3$) and on small scales ($k>0.015$), where noise is high and  signal is low (cf. Fig.~\ref{fig:diff_pk}). The cause of this bias has not been identified. To correct for it, we produced mock spectra covering the range in $P^{noise}$ and $P^{raw}$ observed in the data. While the noise is white, the $k$-dependence of $P^{raw}$ provides  a wide range of relative values of $P^{noise}$ and $P^{raw}$ with each power spectrum. 
We measured  a systematic overestimate of the power spectrum (cf. Fig.~\ref{fig:bias_LKL}), which we  modeled  by $c_0 + c_1\times P^{noise}/P^{raw} + c_2\times  P^{noise}$. We found $c_0=0.999$, $c_1=0.082$, and $c_2=0.007$. This bias is determined from a full analysis (determination of the quasar continuum, correction for spectrograph resolution and for noise); it thus takes  the systematic biases from all the above steps into account. We assign a systematic uncertainty on the resulting power spectrum equal to the 30\% of the correction (cf. Fig.~\ref{fig:systLKL}). 

\begin{figure}[h]
\begin{center}
\epsfig{figure= 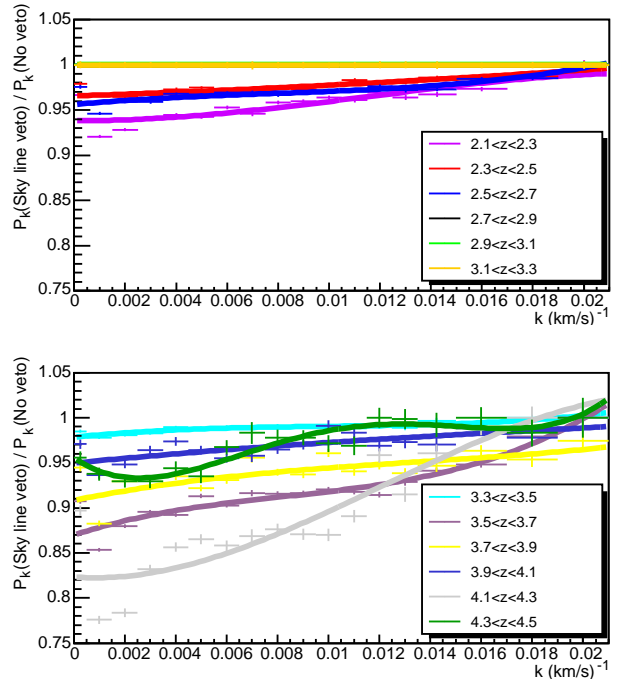,width = \columnwidth} 
\caption[]{\it  Underestimate of the power spectrum due to the masking of the sky emission lines   for the FT approach. The curves are polynomial fits to the measured $k$-dependent bias for each redshift bin. No strong sky lines enter the forest in $2.7<z<3.3$, implying no systematic uncertainty in this redshift range.}
\label{fig:skylines}
\end{center}
\end{figure}

\begin{figure}[h]
\begin{center}
\epsfig{figure= 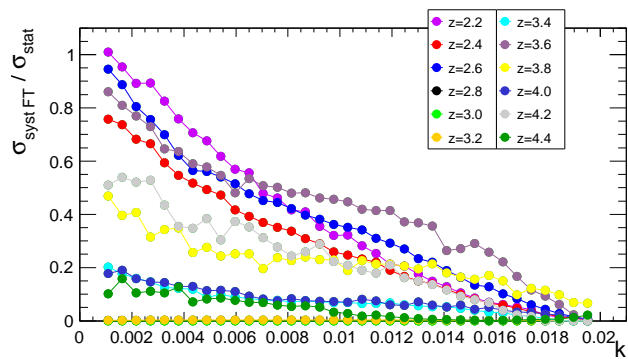,width = \columnwidth} 
\caption[]{\it Systematic uncertainty related to the masking of the sky lines in the FT approach, relative to the statistical uncertainty,  for each redshift bin.}
\label{fig:systFFT}
\end{center}
\end{figure}

The masking of the sky emission lines is implemented in different ways in the two analysis methods. In the likelihood approach, where the relevant pixels are simply omitted, the masking procedure results in no measurable bias. For the FT approach, we estimate the impact of the masking procedure  by applying it on mock spectra that do not include emission from sky lines. The result is illustrated in Fig.~\ref{fig:skylines}. No strong sky line enters the forest for the redshift range $2.7<z<3.3$, which explains why no bias is observed in the corresponding redshift bins. The  largest bias occurs for large-scale modes where most of the  effect is  related to the  relative number of masked pixels in the forest. The effect on small scales is more sensitive to the  distribution and size of the masked regions. The bias tends to decrease with increasing $k$, to become negligible near $k=0.02\;\rm{(km/s)^{-1}}$. It is modeled by  a third-degree  polynomial (except for the  $4.3<z<4.5$ redshift bin where a fourth-degree polynomial is used) that is used to correct the measured power spectrum. 
We assign a systematic uncertainty on the resulting power spectrum equal to the 30\% of the correction. As illustrated in Fig.~\ref{fig:systFFT}, the systematic uncertainty is greater at small $k$, but it remains subdominant  compared to the statistical uncertainty for  all modes. 

Table~\ref{tab:syst_methods} summarizes the sources of bias identified in both analysis methods. The final power spectra are corrected for these under- or overestimations.  As explained above, we infer $k$- and $z$-dependent systematic uncertainties associated with these corrections. Their values are given along with the power spectrum measurements  in Sec.~\ref{sec:results}.
\begin{table}[htdp]
\caption{\it Bias introduced at different steps of the analyses. }
\begin{center}
\begin{tabular}{lcc}
\hline
\hline
 & Fourier transform & Likelihood \\
\hline
QSO continuum & 1.02 & 1.02\\
Spectrograph resolution & $ - $ & $ - $\\
Noise in the data* & $0.98$ & 1.00 to 1.13\\
Masking of sky  lines & 0.82 to 1.00 &  1.00 
\end{tabular}
\end{center}
\it *: the noise-related bias was measured in the Fourier transform  using the  true continuum and is  to be added to the other  biases; for the likelihood, it  includes  systematic effects from all steps. 
\label{tab:syst_methods}
\end{table}

\label{sec:methods_syst}

\subsection{Instrumental uncertainties and associated systematics} 
\label{sec:sys_inst}
The two main sources of instrumental uncertainties are related to the estimate of the noise and the resolution. The techniques to correct these two effects are respectively described in Secs.~\ref{sec:noise} and~\ref{sec:psf}. Here we  present the associated systematics.

\begin{figure}[htbp]
\begin{center}
\epsfig{figure= 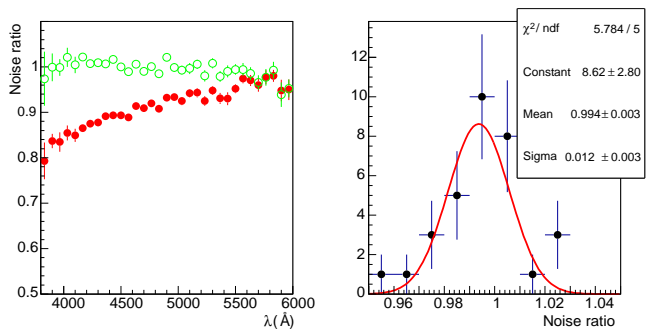,width = \columnwidth} 
\caption[]{\it  Left:  Ratio of pipeline to side-band estimates of the noise as a function of the wavelength without correction (red dots) and when including the correction of Eq.~\ref{eq:corr} (green circles). Right: distribution of the residuals of the noise ratio with including noise correction; the rms of the green distribution, $\sim 1.5\%$, gives an estimate of the uncertainty on the noise correction.} 
\label{fig:PullNoise}
\end{center}
\end{figure}

The power spectrum of the noise is obtained by computing the Fourier transform of a `difference spectrum'  between the individual exposures of a single quasar. In a similar way as  in Fig.~\ref{fig:noise}, we compare the side-band measurement of the noise  (estimated as the flux rms in the $1330<\lambda_{\rm RF}<1380\,$\AA\  side-band of a quasar) either to the pipeline noise or  to our determination of the noise from the difference power spectrum. Using the distribution of quasar redshifts, we show these distributions  as a function of  wavelength in the lefthand plot of Fig.~\ref{fig:PullNoise}: the red curve shows the ratio of the average pipeline noise over the side-band noise; the green curve is the ratio of our estimate of the pixel noise, using the correction factor given in Eq.~\ref{eq:corr}, over the side-band noise. After  correction, the distribution is flat  in wavelength and centered on 1.0, as expected. The righthand plot of Fig.~\ref{fig:PullNoise} shows the distribution of the residuals after correction. Its spread provides an estimate of the remaining uncertainty on the noise. From a  Gaussian fit as shown in righthand plot of Fig.~\ref{fig:PullNoise}, we assign a conservative  $\sim 1.5\%$  systematic error on the noise estimate.

\begin{figure}[htbp]
\begin{center}
\epsfig{figure= 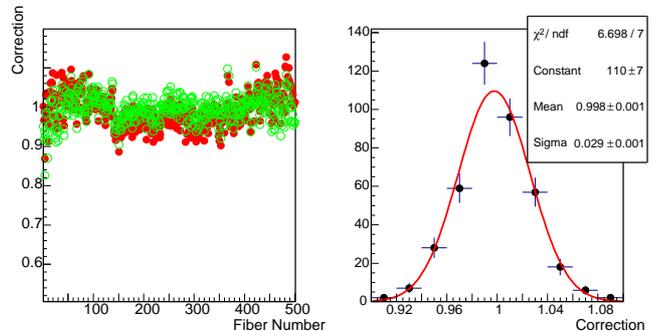,width = \columnwidth} 
\caption[]{\it Left:  discrepancy between pipeline and arc lamp resolution as a function of the fiber number before correction (red dots) and after correction of Sec.~\ref{sec:psf} (green circles). This plot is obtained for the Cd line at $\sim 4800\,${\rm \AA}. Right: distribution of the residuals of the resolution correction; the rms of the green distribution, $\sim 3.0\%$, provides an estimate of the uncertainty on the resolution correction.} 
\label{fig:PullReso}
\end{center}
\end{figure}

We applied a similar method to derive the systematic error related to the resolution. In this case, we plotted the ratio of our resolution measurement  to the  resolution given by the pipeline (red) as a function of the fiber number (see the left plot of Fig.~\ref{fig:PullReso}). In green, the pipeline resolution is corrected by our model of Sec.~\ref{sec:psf}. The rms of the residual distribution with respect to 1.0 yields a value of about $3\%$  for the systematic error on the spectrograph resolution.

\begin{figure}[htbp]
\begin{center}
\epsfig{figure= 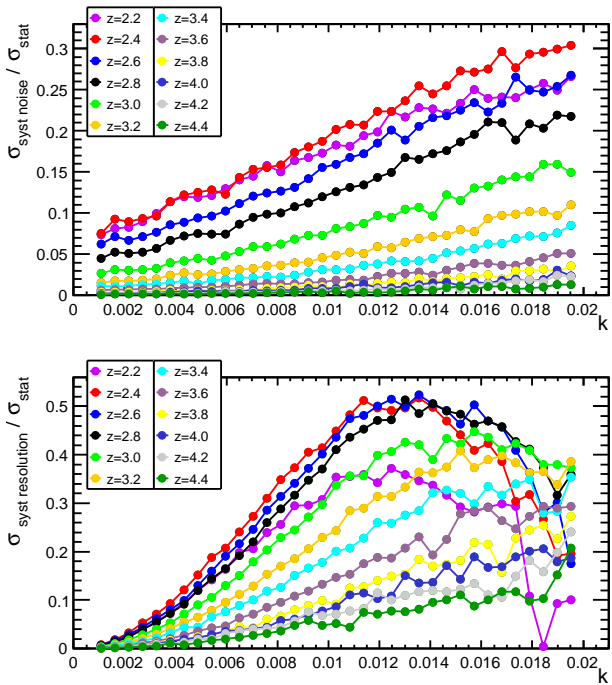,width = \columnwidth} 
\caption[]{\it Systematic uncertainty related to the estimate of the noise (upper plot) or of the spectrograph resolution (bottom plot), relative to the statistical uncertainty,  for each redshift bin.} 
\label{fig:systNR}
\end{center}
\end{figure}

We determine the final impact of each of these two systematic effects using the data. We increase, for instance, our estimate of the noise for all the quasar spectra selected for the data analysis by the observed dispersion of $1.5\%$. We then apply the full procedure to measure the 1D power spectrum $P(k,z)$ with this new estimate of the noise. Finally, for each bin, we compare the new power spectrum,  $P^{\rm new}(k,z)$, to the nominal power spectrum $P^{\rm init}(k,z)$. We define the   systematic error to be   $\sigma^{\rm sys}_P(k,z) \equiv 30\% \times| P^{\rm new}(k,z)-P^{\rm init}(k,z)|$.  This is a  conservative approach since we here consider a systematic effect acting  in the {\em same} direction for all the quasars. The impact on the power spectrum of these systematic uncertainties are illustrated in Fig.~\ref{fig:systNR}. The systematic on the noise estimate is the largest for low-redshift bins; with the cut we have applied on the spectrum S/N,   its contribution is at most of 70\% of $\sigma_{\rm stat}$. The systematic on the resolution estimate becomes dominant, over all other sources of uncertainties, on small scales  for $z<3.0$. The stringent cut we have applied on the  mean resolution in the forest ($\overline{R} <85$~km/s), however, has limited this  uncertainty to be at most $1.2\,\sigma_{\rm stat}$, instead of  $\sim5.5\,\sigma_{\rm stat}$ that  would have existed in the absence of such a cut.

 \label{sec:instrument_syst}

%%%%%%%

\section{Power spectrum measurement}\label{sec:results}
We apply the methods presented previously to the BOSS data and measure the flux power spectrum in the Ly$\alpha$ forest region. It contains two components: the signal arising from  \ion{H}{i} absorption, and the background due to  absorption by all  species other than atomic hydrogen (hereafter `metals'). In this section, we explain how we separate each contribution and  conclude by summarizing the obtained results.

Absorption at an observed wavelength $\lambda$ receives
contributions from any atomic species, $i$, absorbing at wavelength $\lambda_i$, if
the absorption redshift
$z_i+1=\lambda/\lambda_i$ satisfies $z_i<z_{\rm qso}$.  
We want to subtract  the  background from metals.  To do this, we use two methods
that work  for species with $\lambda_i>\lambda_{Ly\alpha}$
and  $\lambda_i\sim\lambda_{Ly\alpha}$.  

For the first case,  the wavelength of the metal line  is far from  Ly$\alpha$. If its absorption falls in the Ly$\alpha$ forest  of a quasar, then the Ly$\alpha$ absorption from the same redshift absorber is outside (bluer than) the forest wavelength range. It  therefore  presents no correlation with the Ly$\alpha$ absorption. The
 summed absorption at $\lambda$
due to all such species can be determined
by studying absorption at $\lambda$ in quasars with
$z_{\rm qso}+1<\lambda/\lambda_{Ly\alpha}$ for which Ly$\alpha$ absorption
makes no contribution. The subtraction of the background for this first case is described in Sec.~\ref{sec:metals}. 
For the second case,  atomic hydrogen and the metal species produce correlated absorption within the Ly$\alpha$ forest~\citep{bib:pieri10}. The  1D correlation
function will have a peak at wavelength separations corresponding to
hydrogen and metallic  absorption at the \emph{same} redshift:
$\Delta\lambda/\lambda=1-\lambda_i/\lambda_{Ly\alpha}$. The main contribution in this second case comes from  \ion{Si}{iii}. The strategy adopted to subtract this second category of background is described in Sec.~\ref{sec:si3}. 
Then, in Sec.~\ref{sec:resultsummary}, we present the final results in such a way that the reader can access directly the signal power spectrum and the different contaminating components.

\subsection{Uncorrelated background subtraction}
\label{sec:metals}

The uncorrelated background due to metal absorption in the Ly$\alpha$ forest  cannot be estimated directly from the power spectrum measured in this region. We address this issue by estimating the background components in sidebands located at longer wavelengths than the  Ly$\alpha$ forest region. We measure the power spectrum in these sidebands and  subtract it from the Ly$\alpha$ power spectrum measured in the same gas redshift range. This method is purely statistical;  we use different quasars to compute  the Ly$\alpha$ forest and the metal power spectra for a given redshift bin. This approach is inspired by the method described in~\cite{bib:mcdonald06}. However, our approach is simpler and more robust because it  only relies   on control samples and does not require any modeling. 

In practice, we define two sidebands that correspond, in the quasar rest frame, to the wavelength ranges  $ 1270 < \lambda_{\rm RF}< 1380 \,$\AA\ and $ 1410 < \lambda_{\rm RF}< 1520\,$\AA. The  power spectrum measured in the first sideband includes the contribution from all metals with  $\lambda_{\rm RF}>1380\,$\AA, including absorption from \ion{Si}{iv} and \ion{C}{iv}. The second sideband also includes  \ion{C}{iv}  but excludes the \ion{Si}{iv} absorption.  For our analysis, we use the first sideband ($ 1270 < \lambda_{\rm RF}< 1380\,$\AA) to subtract the metal contribution in the power spectrum,  and measurement in the second sideband constitutes an important consistency check. 

 We  determine the metal power spectrum in  the same observed wavelength range as the Ly$\alpha$ forest power spectrum from which it is being subtracted. For instance, for the first redshift bin,  $2.1<z<2.3$, we measure the power spectrum in the first sideband,  corresponding to  $ 3650 < \lambda< 4011\,$\AA, i.e., using quasars with a redshift $z\sim1.9$. Quasars in a given redshift window have their two sidebands corresponding to fixed observed wavelength windows, which in turn match a specific redshift window of Ly$\alpha$ forest. 

\begin{figure}[htbp]
\begin{center}
\epsfig{figure= 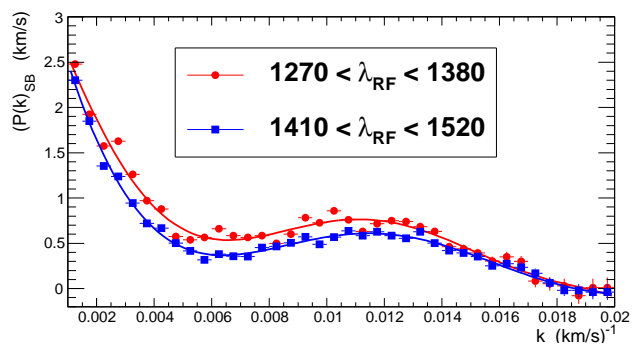,width = \columnwidth} 
\caption[]{\it Power spectrum $P_{SB}(k)$ computed for sideband regions above the Ly$\alpha$ forest region. The red dots  and the blue squares are  for the two sidebands defined in the rest frame by $ 1270 < \lambda_{\rm RF}< 1380\,${\rm \AA} and $ 1410 < \lambda_{\rm RF}< 1520\,${\rm \AA} respectively. Each power spectrum is fitted with a  sixth-degree polynomial.} 
\label{fig:metalsAll}
\end{center}
\end{figure}

\begin{figure}[htbp]
\begin{center}
\epsfig{figure= 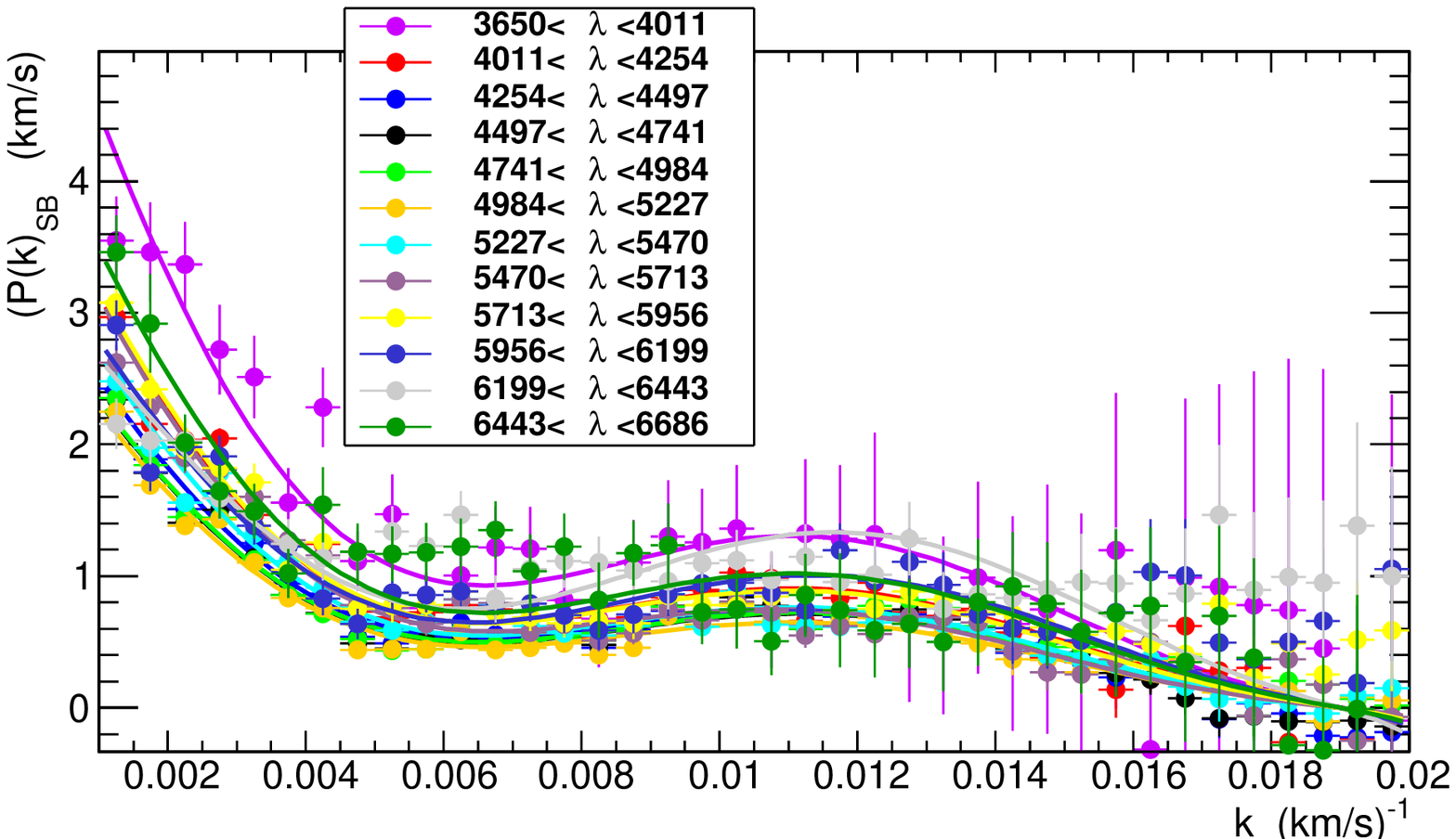,width = \columnwidth} 
\epsfig{figure= 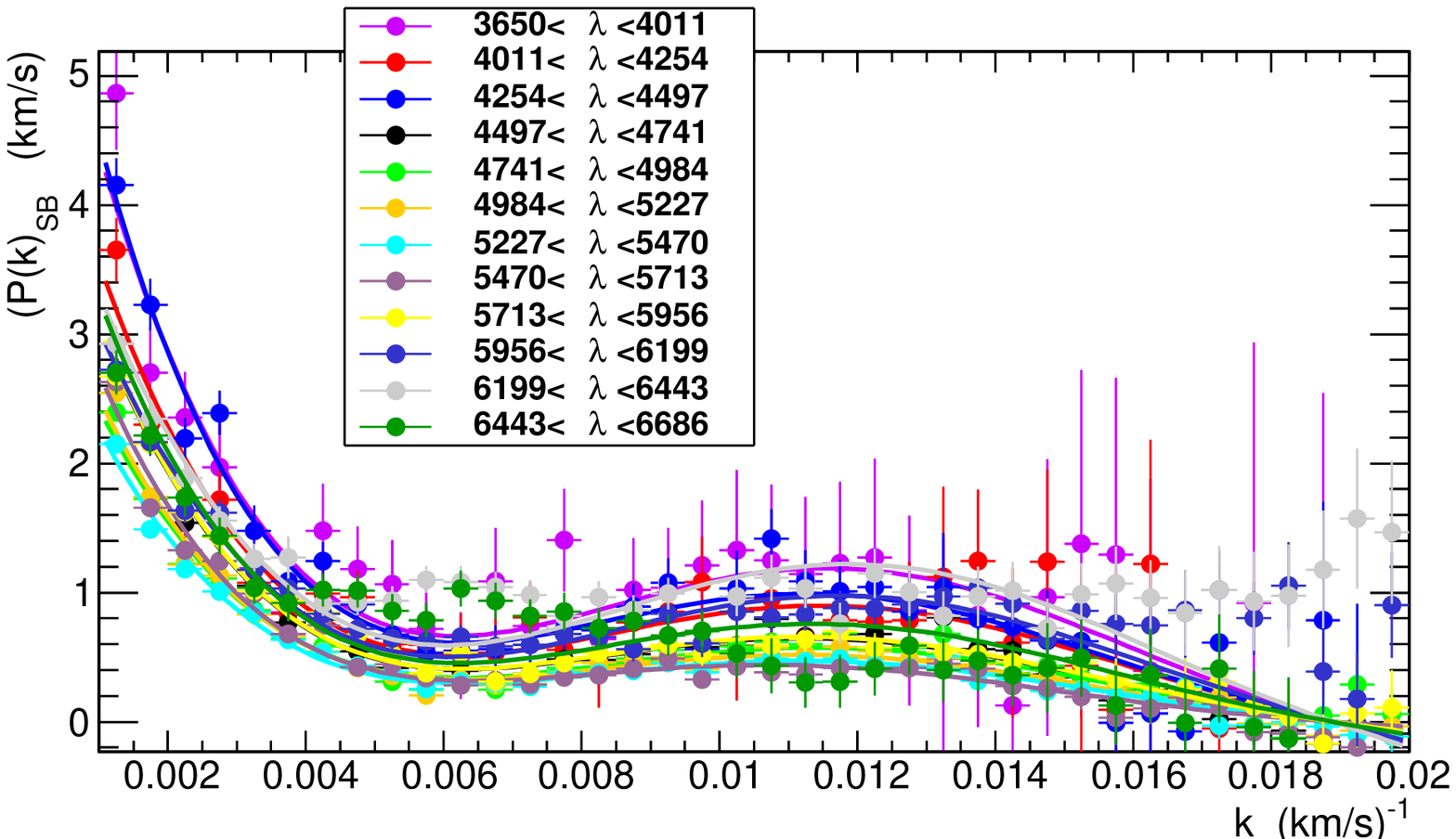,width = \columnwidth} 
\caption[]{\it Power spectrum $P(k)$ computed for sideband regions above the Ly$\alpha$ forest region for different $\lambda$ windows. Each $\lambda$ region corresponds to one redshift bin.
The top and bottom plots correspond respectively to the two sidebands defined in the rest frame by $ 1270 < \lambda_{\rm RF}< 1380\,${\rm \AA} and $ 1410 < \lambda_{\rm RF}< 1520\,${\rm \AA}. Each power spectrum is fitted by the product of the sixth-degree polynomial obtained in Fig.~\ref{fig:metalsAll} and a first-degree polynomial in which the 2 parameters are free.} 
\label{fig:metalsbin}
\end{center}
\end{figure}

The power spectra $P_{SB}(k)$ shown in Fig.~\ref{fig:metalsAll}  are obtained with $\sim 40,000$ quasars with redshift in the range $1.7<z<4.0$, passing similar quality cuts as the quasars for the Ly$\alpha$ forest analysis.  The shapes of $P_{SB}(k)$ are similar for the two sidebands. As expected, for the second sideband, corresponding to $ 1410 < \lambda_{\rm RF}< 1520\,$\AA, which excludes \ion{Si}{iv}, the magnitude of  $P_{SB}(k)$  is smaller.  We fit the distribution $P_{SB}(k)$ with a sixth-degree polynomial. We  use this fitted function as a template to parametrize the $P_{SB}(k)$ measured for each wavelength window (see  Fig.~\ref{fig:metalsbin}). 

As the shape and the magnitude of the power spectrum vary from one  wavelength window to another, we have parameterized this as the product of  the fixed shape obtained in Fig~\ref{fig:metalsAll}, with a variable first-degree polynomial, with two free parameters that are different for each wavelength window. This adequately fits the measured power in all the wavelength windows (see Fig.~\ref{fig:metalsbin}). From these parametric functions, we extract the value of the power spectrum $P_{SB}(k)$  for each $k$ and for each Ly$\alpha$ redshift window. 

The statistical uncertainty on $P_{SB}$  is  strongest where we have the lowest number of quasars to measure the metal contribution. This occurs in the $z\sim 2.2$ redshift bin for which we only have $\sim400$ quasars (at $z_{\rm qso}\sim 1.7$) instead of  about 4000 on average for the other bins. For $z\sim 2.2$,  the uncertainty on the metal correction, derived from the statistical precision on  the first-degree polynomial fit, is around 10\%. 

An uncertainty on our metal correction will have the strongest impact relative to the measured $P(k)$ in the Ly$\alpha$ region when the absolute $P(k)$ has the lowest value. This again occurs for  $z\sim 2.2$, which therefore constitutes the worst case both in terms of statistical uncertainty and relative level of the correction. Even in this worst case, the  metal power spectrum is less than 10\% of the Ly$\alpha$ power spectrum. The  uncertainty of the metal correction is therefore less than 1\% of the  Ly$\alpha$ $P(k)$  across our whole sample.

\subsection{\ion{Si}{iii} cross-correlation}
\label{sec:si3}

The correlated background due to absorption by Ly$\alpha$ and \ion{Si}{iii} from the same gas cloud along the quasar line of sight can be estimated directly in the power spectrum. Since \ion{Si}{iii} absorbs at $\lambda = 1206.50\,$\AA, it appears in the data auto-correlation function $\xi_{\rm tot}(v) = \langle \delta(x) \delta (x+v)\rangle$ as a bump at $\Delta v = 2271$~km/s, and in the power spectrum as wiggles with peak separations of $\Delta  k=2\pi/\Delta v = 0.0028\;(\rm km/s)^{-1}$. Following the approach suggested by \citet{bib:mcdonald06}, we model the \ion{Si}{iii} structure as being equal to that of the Ly$\alpha$ forest up to an overall normalization: $\delta_{\rm tot} = \delta(v) + a\delta(v+\Delta v)$ where $\delta (v)$ is only for Ly$\alpha$. The corresponding correlation function is
\begin{equation}
\xi_{\rm tot}(v) = (1+a^2)\,\xi(v) + a\,\xi(v+\Delta v) + a\,\xi(v-\Delta v)
\end{equation}
and the corresponding power spectrum
\begin{equation}
P_{\rm tot} = (1+a^2)\,P(k) + 2\,a\cos(\Delta v\,k)P(k)\;,
\label{eq:siIII}
\end{equation}
where $\xi(v)$ and $P(k)$ are for Ly$\alpha$-Ly$\alpha$ correlations. 
We clearly detect, in the power spectrum, the oscillatory pattern due to the \ion{Si}{iii}-Ly$\alpha$ cross correlation (cf. Fig.~\ref{fig:PkWithSi3}), or equivalently a peak near $\Delta\lambda=9.2\,$\AA\ in the correlation function.
%The correlation function is shown in Fig.~\ref{fig:siIII_correl}.  A clear peak appears near $\Delta\lambda=9.2\,$\AA, corresponding to the \ion{Si}{iii}-Ly$\alpha$ cross correlation. 
We do not observe any other significant metal features seen in \citet{bib:pieri10}, such as \ion{Si}{ii} lines (at 22.4~\AA\ and 25.3~\AA) or \ion{N}{v} lines (at 23.2~\AA\ and 27.1\AA). However, some weak contribution may be present from metals where they do not produce signal distinct from each other or from the greater Ly$\alpha$ signal.

The measured normalization evolves with redshift roughly as $a(z) = f_ {\rm{Si\,III}}/ (1-\bar{F}(z))$, where $\bar{F}(z)$ is the mean transmitted fraction defined in Sec.~\ref{sec:cont}. With a simple fit, we find a normalization factor $f_ {\rm{Si\,III}} = 0.008\pm 0.001$, similar to the value $f \sim 0.011$ measured by~\citet{bib:mcdonald06} on a sample of 3000 SDSS quasars, and in agreement with the value derived in Sec.~\ref{sec:cosmo} from a completely independent fit to a cosmological model.

%\begin{figure}[htbp]
%\begin{center}
%\epsfig{figure= plots/siIII_correl.eps,width = \columnwidth} 
%\caption[]{\it Normalized correlation function $\xi_{\rm tot}(\lambda)/ \xi_{\rm tot}(0)$ in the Ly$\alpha$ forest region. A clear peak is visible at $\Delta\lambda=9\,${\rm \AA}  corresponding to the Ly$\alpha$-\ion{Si}{iii} cross correlation.}
%\label{fig:siIII_correl}
%\end{center}
%\end{figure}

\subsection{Summary of experimental results} \label{sec:resultsummary}

\begin{figure}[htbp]
\begin{center}
\epsfig{figure= 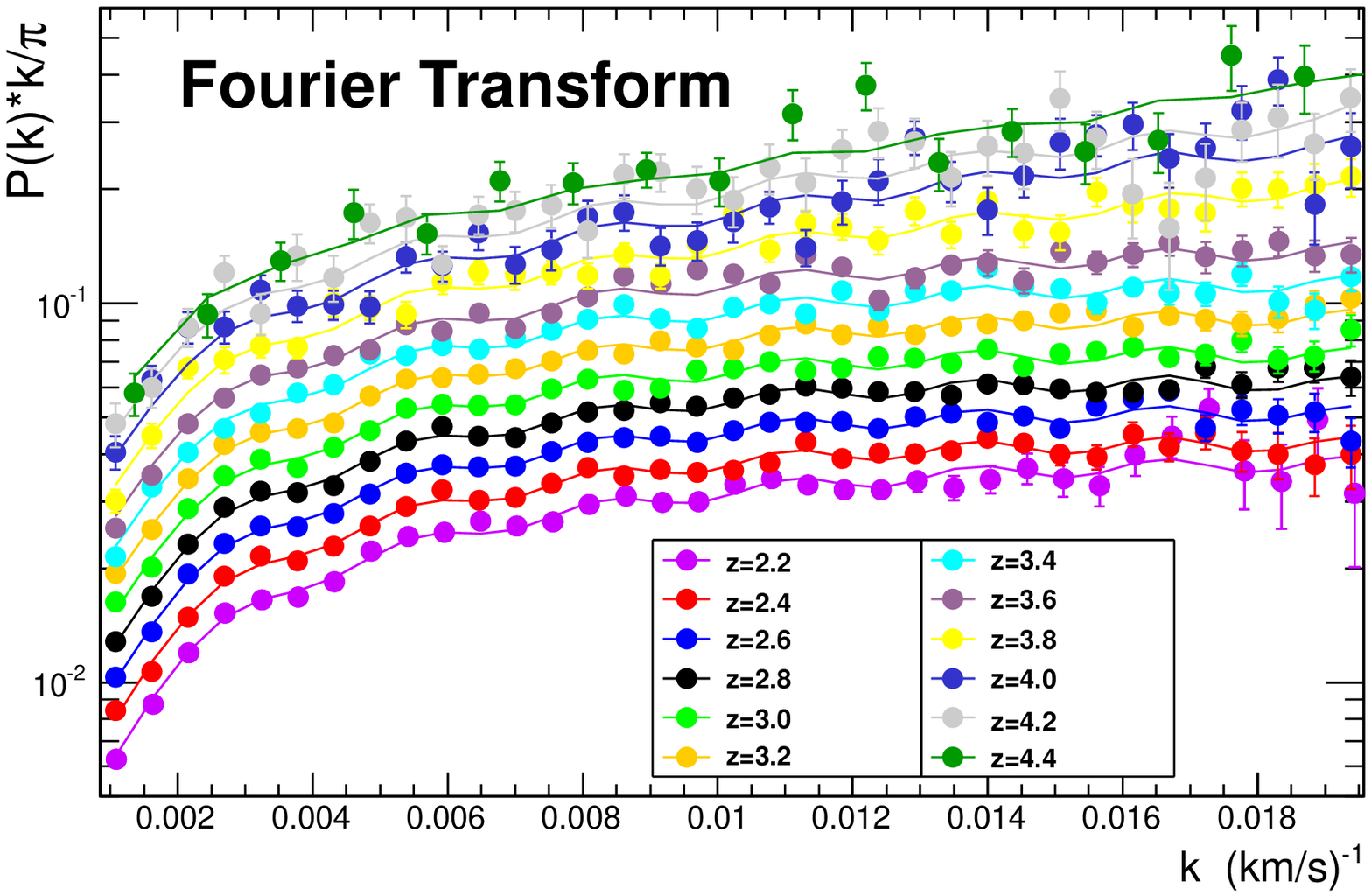,width = \columnwidth} 
\epsfig{figure= 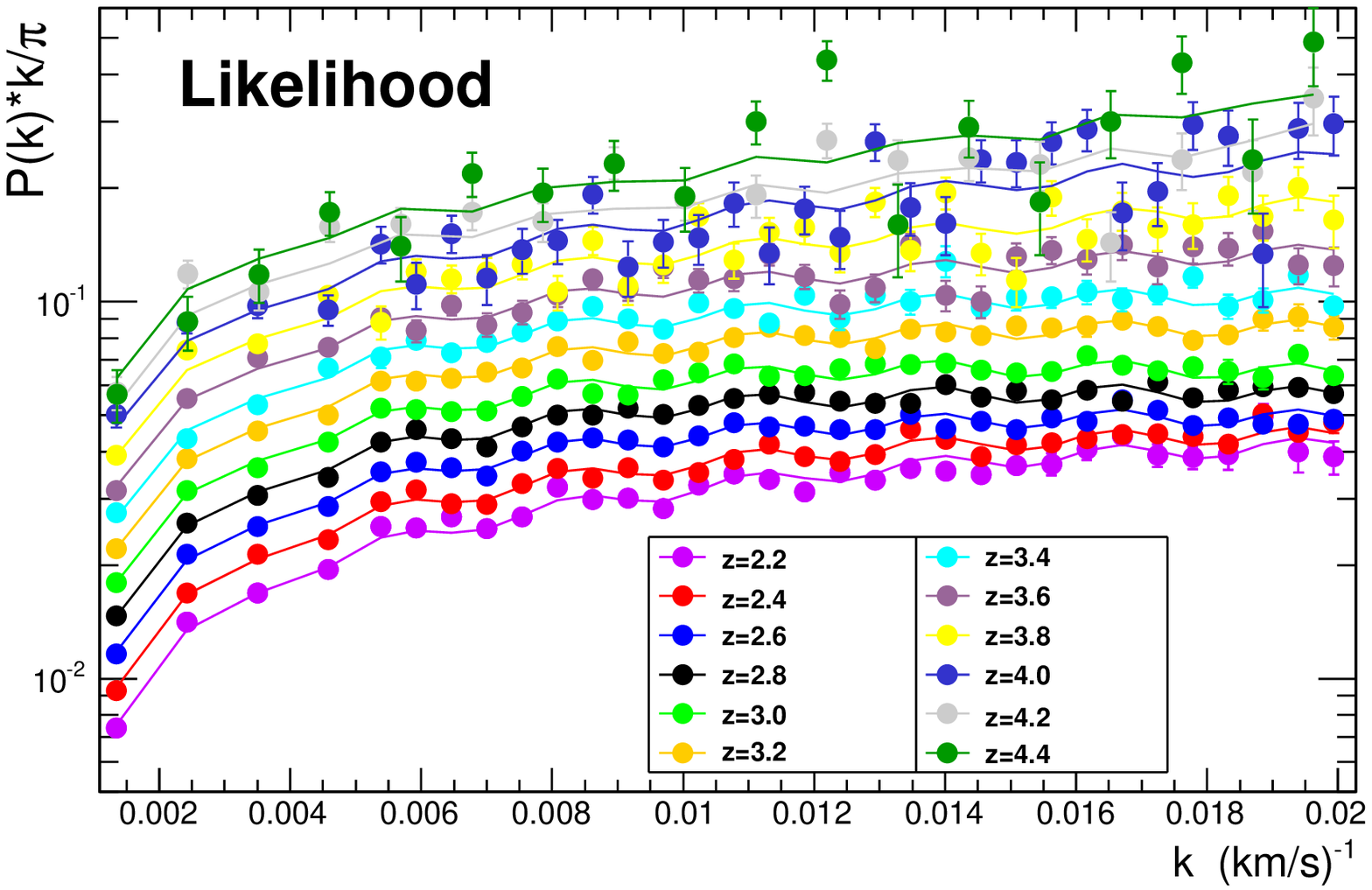,width = \columnwidth} 
\caption[]{\it One-dimensional Ly$\alpha$ forest power spectrum obtained with the Fourier transform method (top plot) and the likelihood method (bottom plot). The metal contribution estimated in Sec.~\ref{sec:metals}, is subtracted. The power spectrum is fitted with the empirical function of Eq.\ref{eq:PkFunc}. } 
\label{fig:PkWithSi3}
\end{center}
\end{figure}

\begin{figure}[htbp]
\begin{center}
\epsfig{figure= 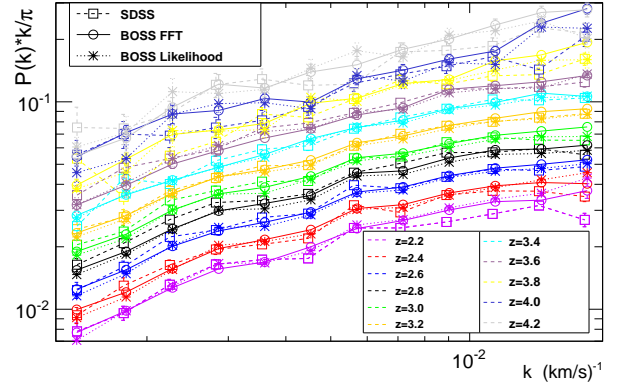,width = \columnwidth} 
\caption[]{\it Comparison  of the 1D Ly$\alpha$ forest power spectrum obtained  in BOSS  and in SDSS (see~\cite{bib:mcdonald06})  over the redshift range,  $z= [2.1-4.3]$. For BOSS, we show the results for the  two methods, Fourier transform and likelihood, and we use the same $k$ bins as in~\cite{bib:mcdonald06}.} 
\label{fig:PkComparison}
\end{center}
\end{figure}

Figure~\ref{fig:PkWithSi3} shows the one-dimensional Ly$\alpha$ forest power spectrum obtained with  the Fourier transform  and the likelihood method. Figure~\ref{fig:PkComparison} demonstrates good agreement between the methods, although they are quite different in the treatment of the sky line masking, the noise subtraction, and the resolution correction.  Moreover, the agreement with the previous SDSS measurements~\citep{bib:mcdonald06} is also remarkable. The only significant discrepancy between SDSS and BOSS is observed for the low $z$ and high $k$ region where  the noise subtraction is  difficult. The uncertainty in this region is covered either by the use of the systematics errors  given in Sec.~\ref{sec:sys_inst} or by introducing  nuisance parameters. The latter option is the one we have chosen for the cosmological interpretation of our results (see Sec.~\ref{sec:nuisance}).

To compare the measured power spectrum for SDSS and BOSS, and also to compare the results of the Fourier transform  and the likelihood methods in a quantitively way, we define an empirical function $P^{\rm emp}$ with which we fit each power spectrum distribution.  This function, written in Eq.~\ref{eq:PkFunc}, has five free physical parameters: an amplitude $A_F$ corresponding to the amplitude of the power spectrum at the pivot mode $k_0$ and redshift $z_0$, a slope $n_F=d\ln P/d\ln k|_{(k_0, z_0)}$, a curvature $\alpha_F = d\ln n_F/d\ln k|_{(k_0, z_0)}$, and two parameters, $B_F$ and $\beta_F$, that model the redshift evolution of the power spectrum.   In addition, we introduce  nuisance parameters to take  the correlation between  \ion{H}{i} and \ion{Si}{iii} into account (parameter $a$ in Eq.~\ref{eq:PkFunc}), and the imperfection of our resolution and noise models.  We choose a pivot point in the middle of our measurements, $k_0=0.009$~(km/s)$^{-1}$ and $z_0=3.0$.  The results of the fits are summarized in Table~\ref{tab:ResEmpFit}. The agreement between the different methods and datasets is  good. All five parameters are within 1 or 2$\sigma$ of one another.

\begin{eqnarray}
\frac{k\;P^{\rm emp} (k,z)}{\pi} & =  & A_F \times  \biggl(  \frac{k}{k_0}\biggr)^{3+n_F
+\alpha_F \ln(\frac{k}{k_0})  + \beta_F \ln(\frac{1+z}{1+z_0}) }
 \nonumber\\
 & &	\times \biggl( \frac{1+z}{1+z_0} \biggr)^{B_F}  \nonumber\\
 & & \times (1+a^2 + 2\,a\cos(\Delta v\,k))
\label{eq:PkFunc}
\end{eqnarray}

\begin{table}[htdp]
\caption{\it Results of the fit by the empirical function $P^{emp} (k,z)$ (see definition in Eq.~\ref{eq:PkFunc})  of the SDSS and BOSS datasets over the redshift range,  $z= [2.1-4.3]$. These five parameters should not be used for any quantitative science since the $\chi^2$ remain $\sim 1.4$ even after adding nuisance parameters in the fit.}
\begin{center}
\begin{tabular}{cccc}
\hline
\hline
Parameter & SDSS & BOSS & BOSS  \\
& & FT & likelihood \\
\hline
$A_F$ & $0.062\pm0.002$ &  $0.067\pm0.001$ &  $0.064\pm0.001$\\
$n_F$ & $-2.64\pm0.04$  & $-2.50\pm0.02$ &  $-2.55\pm0.02$ \\
$\alpha_F$ &  $-0.13\pm0.02$ & $-0.08\pm0.01$  &  $-0.10\pm0.01$ \\
$B_F$ &  $3.3\pm0.14$ &  $3.36\pm0.06$  &  $3.55\pm0.07$\\
$\beta_F$ & $-0.28\pm0.09$ & $-0.29\pm0.04$ &  $-0.28\pm0.05$ \\
\end{tabular}
\end{center}
\label{tab:ResEmpFit}
\end{table}

\begin{figure}[htbp]
\begin{center}
\epsfig{figure= 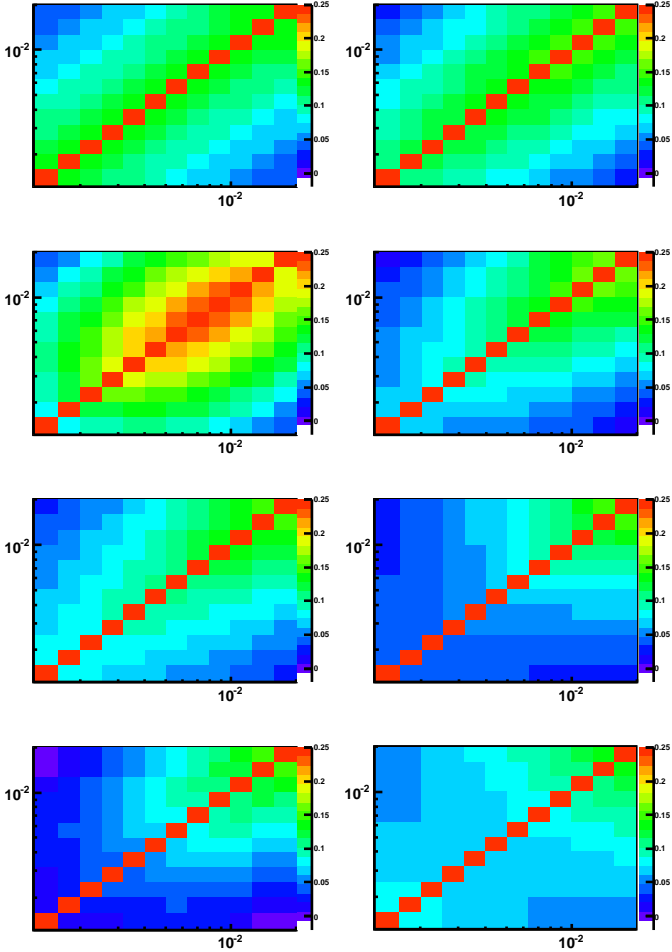,width = \columnwidth} 
\caption[]{\it Correlation matrices  between the different $k$-modes for the first  8 redshift bins ($z=[2.1-3.7]$), smoothed by 2D second-degree polynomials, for the FT method. The color range is identical in all 8 plots, with red for all values above 0.25.} 
\label{fig:correl_smooth}
\end{center}
\end{figure}

%%%%%% To remove for archive submission
Tables~\ref{tab:Resultsfft} and~\ref{tab:Resultslkl}  summarize,  for each redshift bin, the results for the  1D Ly$\alpha$ forest power spectrum. They are available in their entirety in the online edition.\footnote{Online edition and full tables available at the CDS via anonymous ftp to cdsarc.u-strasbg.fr (130.79.128.5), or via http://cdsweb.u-strasbg.fr/cgi-bin/qcat?J/A+A/.}  The different components ($P_ {1D}$,  $P^{noise}$ and $P^{metals}$) are given in these tables.  In  $P^{metals}$  we  consider only the uncorrelated background computed in Sec.~\ref{sec:metals}. The last two columns   represent the statistical and systematical uncertainty on  $P_ {1D}$. We added in quadrature all the systematic uncertainties studied in Sec.~\ref{sec:sys_inst}.
The correlation matrices are illustrated in Fig.~\ref{fig:correl_smooth} for the first eight redshift bins that are used in Sec.~\ref{sec:cosmo} for the cosmological interpretation. The maximum correlation is at the level of $\sim 20\%$ for neighboring $k-$modes, and the correlation rapidly drops to $<10\%$. 
%%%%%%

\begin{table}[htdp]
\caption{\it $P_ {1D}$  results  obtained with  the FT method for each $k$ and $z$ bin (top table), and correlation matrices between $k$ bins for each $z$ bin (following tables). These tables are available in their entirety in the electronic edition of the journal. A portion is shown here for guidance
regarding their form and content. Units of $k$ are $\rm(km/s)^{-1}$, power spectra have units $\rm{km/s}$. }
\begin{center}
\begin{tabular}{ccccccc}
\hline
\hline
$z$ & $k$ &  $P_ {1D}$ & $P^{noise}$  & $P^{metals}$ & $\sigma^{\rm stat}_P$ & $\sigma^{\rm sys}_P$ \\
\hline
2.2 & 0.00108 &18.15&0.52 &7.87&4.41 &0.53\\
2.2 &0.00163&16.83&0.47 &7.86&3.73& 0.45\\
...
\end{tabular}
\vskip .2cm
\begin{tabular}{ccccc}
\hline
\hline
$(z=2.2)$, $k$&0.00108&0.00163& 0.00217&...\\
\hline
0.00108&1&0.112&0.108&...\\
0.00163&0.112&1&0.106&...\\
0.00217&0.108& 0.106&1 &...\\
...
\end{tabular}
\end{center}

\label{tab:Resultsfft}
\end{table}

\begin{table}[htdp]
\caption{\it Same as table~\ref{tab:Resultsfft}, for the likelihood method.}
\begin{center}
\begin{tabular}{ccccccc}
\hline
\hline
z & k &  $P_ {1D}$ & $P^{noise}$  & $P^{metals}$ & $\sigma^{\rm stat}_P $  $\sigma^{\rm sys}_P$ \\
\hline
2.2 & 0.00135 &17.21&0.50 &7.88&4.07 &0.47\\
2.2 &0.00242&18.24&0.37 &8.11&2.83& 0.48\\
...
\end{tabular}
\vskip .2cm
\begin{tabular}{ccccc}
\hline
\hline
$(z=2.2)$, $k$&0.00135&0.00242& 0.00350&...\\
\hline
0.00135&1&-0.26&-0.01&...\\
0.00163&-0.26&1&-0.19&...\\
0.00350&-0.01&-0.19&1 &...\\
...
\end{tabular}
\end{center}

\label{tab:Resultslkl}
\end{table}

%%%%%%%

\section{Cosmological constraints}\label{sec:cosmo}
In this paper, we present a preliminary cosmological interpretation of our results. Our intention is to demonstrate the improvements in our measurements over the previous publication of SDSS presented in~\cite{bib:mcdonald06}. We  use an approach developed by~\cite{bib:viel06}. This method is well adapted to the statistical accuracy of  SDSS  and is sufficient to give some results on the two  cosmological parameters ($\sigma_8$,$n_s$) from our measurement of the 1D power spectrum. 

In this interpretation, we only  use the first eight redshift bins (i.e. $z= [2.1-3.7]$) as  recommended in~\cite{bib:viel06}. Moreover, to break the degeneracies between the cosmological parameters, we use a constraint on $H_0$ that encompasses the measurements of \cite{riess11} and of \cite{bib:planck13}. This section should be seen as an exploration of potential cosmology from  the 1D power spectrum. 

As was shown in the previous section, the FT and the likelihood methods yield compatible results. We therefore restrict the cosmological fits of this section  to the power spectrum obtained with the Fourier transform only.

\subsection{Simulations}
We computed the constraints from our measurements of the flux  power spectrum following~\cite{bib:viel06} and~\cite{bib:viel09} where all the details can be found. We only give  a brief summary here. 

They used a grid of full hydrodynamical simulations run with the Tree-smoothed particle hydrodynamics code \textsc{gadget-2} \cite{bib:springel05} to model the nonlinear relation between the flux and the matter power spectra. The flux power spectrum was calculated using a second-order Taylor expansion  (without cross terms). They ran several hydrodynamical simulations around the best estimated value for each cosmological and astrophysical parameter  to compute the derivatives required for the Taylor expansion. The central cosmology used in the simulation grid is a  $\Lambda$CDM model with 
$\sigma_8=0.85$, $n_s=0.95$, $\Omega_m = 0.26$, and $H_0=72$~km/s/Mpc.
The cosmological parameters are close to the values obtained by WMAP~\citep{bib:hinshaw13}, and the predicted flux statistics are derived by expanding around a model with $\gamma\sim1$.  This  parameter  defines the density-temperature relation of the IGM by the approximate relation $T = T_0(1+\delta)^{\gamma-1}$.

More complex astrophysical effects on statistics of the Ly$\alpha$ flux have been discussed in \citet{bib:viel13}, where it has been explicitly shown that the overall distribution of the gas in the $T-\rho$ plane can affect the flux power in a redshift- and scale-dependent way for different physical models that include active galactic nuclei feedback or galactic winds. The effect is of similar  magnitude as the statistical uncertainty of the data. 
A robust and conservative analysis of the 1D power should properly
take  the impact that astrophysics has on the final observable into account, and model this 
to a high level of precision. 
These effects have not been addressed in this work since the  simulations used do not  allow such refinements. It is however unlikely that the scale and redshift dependence of  feedback models on the flux power is similar to that induced by cosmological parameters.

The standard simulations corresponded to a box of length $L= 60 ~h^{-1}$~Mpc with 2$\times400^3$ (gas + dark matter) particles. They are corrected for  box size effects using simulations with the same number of particles but $L= 120~h^{-1}$~Mpc, and for resolution effects using simulations with $L=60~ h^{-1}$~Mpc  and 2$\times512^3$ particles. 

\subsection{Parameters}
\label{sec:nuisance}
In the modeling of the likelihood, we have introduced four categories of parameters that are floated in the maximization of the likelihood. The first category describes the cosmological model in the simplest case of $\Lambda$CDM assuming a flat Universe with a zero neutrino mass. The second category models the astrophysics within the IGM and the relation between temperature and density for the gas.  The  purpose of the third category is  to describe the imperfections of our measurement of the 1D power spectrum. By fitting the parameters of the latter category, we improve significantly the goodness of the fit but  we reduce the sensitivity on the other parameters.  Finally, a last parameter allows for a residual contamination from damped Ly$\alpha$ systems in our selected sample of quasar lines of sight. 
\begin{itemize}
\item {\bf Cosmological parameters:} Our $\Lambda$CDM cosmology is described by the fluctuation amplitude of the matter power spectrum $\sigma_8$, the spectral index of primordial density fluctuations, $n_s$, the matter density $\Omega_m$, and the Hubble constant $H_0$.
\item {\bf Astrophysical parameters:} Two parameters describe the effective optical depth assuming a power law evolution, $\tau_{\rm eff}(z) =  \tau_{{\rm eff,}z=3}^A\times [(1+z)/4 ]^{\tau_{\rm eff}^S}$. The evolution with redshift of $\gamma$ and $T_0$ are also modeled with power laws. For the temperature, we have two parameters: $T^A_{0, z=3}$, the temperature  at  $z=3$,  and the slope (or exponent) $T^S_{0}$. Similarly for $\gamma$, we have two parameters: $\gamma^A_{ z=3}$, the value  at  $z=3$,  and the slope $\gamma^S$. To account for the effect of the correlated \ion{Si}{iii}  absorption, we  introduce a multiplicative  term, $1+a^2+2a\cos(vk)$ with $a = f_ {\rm{Si\,III}}/(1-{\bar F}(z))$ as  in Sec.~\ref{sec:si3} following the suggestion  of~\cite{bib:mcdonald06}. The parameter $f_ {\rm{Si\,III}}$ is free in the fit, and $v$ is fixed at 2271~km/s.
\item {\bf Nuisance parameters:}  We  take  the imperfection of our resolution model into account by floating one multiplicative term. We allow for imperfection in our noise estimate by floating eight additive terms (one for each redshift bin). 
\item {\bf Damped Ly$\alpha$ system}: In~\cite{bib:mcdonald05}, the effect of the DLA was modeled on the power spectrum. We  included this correction using their $k$ dependence. We have the possibility of fitting the amplitude of this correction with the parameter $A_{\rm damp}$. 
\end{itemize}

\subsection{Fit to the SDSS and BOSS data}

For a given cosmological model defined by the $n$ cosmological, astrophysical and nuisance parameters $\Theta=(\theta_{1},\ldots,\theta_{n})$, and for a data set of power spectra $P(k_i,z_j)$ measured with Gaussian experimental errors $\sigma_{i,j}$, the likelihood function can be written as
\[
{\cal L}\bigl(P,\sigma;\Theta\bigr) = \prod_{i,j}\frac{1}{\sqrt{2\pi}\sigma_{i,j}}
exp\biggl(
-\frac{[P(k_i,z_j) - P^{th}(k_i,z_j) ]^2}{2\sigma_{i,j}^2}
\biggr)\nonumber
\]
where $ P^{th}(k_i,z_j)$ is the  predicted value of the power spectrum for the bin $k_i$ and redshift $z_j$.

In the rest of this paper, we adopt a $\chi^2$ notation, which means
that the following quantity is minimized:

\begin{equation}
\chi^2(P,\sigma;\Theta) = -2 \ln ({\cal L}(P,\sigma;\Theta)) + \chi^2_{\rm ext}(H_0) \,.
\label{eq:chi2}
\end{equation}

The second term of Eq.~\ref{eq:chi2} represents the external constraint on $H_0$. Because the value obtained from HST observations ($H_0^{\rm HST}=73.8 \pm 2.4$) given in~\cite{riess11} and the value from Planck ($H_0^{\rm Planck}=67.4 \pm 1.4$) given in~\cite{bib:planck13} differ by over $2\sigma$, we write this $H_0$ constraint so as to give equal weight to both measurements:\\
\begin{tabular}{rl}
\\
$\chi^2_{\rm ext}(H_0)=\left\{  
\begin{array}{lcl} 
\frac{(H_0-H_0^{\rm Planck})^2}{\sigma_{\rm Planck}^2}  &{\rm if} &  H_0<H_0^{\rm Planck}\\
0  &{\rm if} & H_0^{\rm Planck}<H_0< H_0^{\rm HST}\\
\frac{(H_0-H_0^{\rm HST})^2}{\sigma_{\rm HST}^2} &{\rm if} & H_0^{\rm HST}<H_0\;.
\end{array}
\right. $\\
\\
\end{tabular}

\begin{figure}[htbp]
\begin{center}
\epsfig{figure= 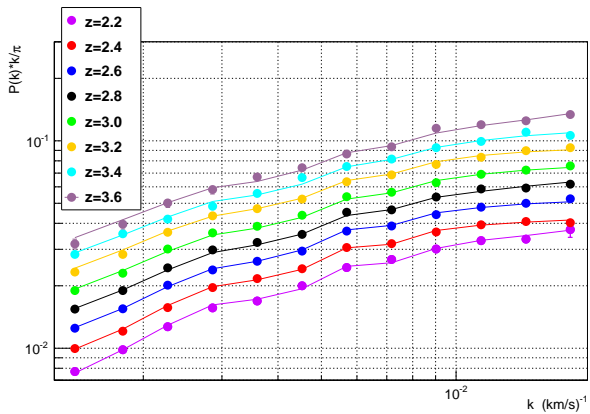,width = \columnwidth} 
\caption[]{\it  
Fit of the power spectrum measured with BOSS in the range $z= [2.1-3.7]$;  the $z$ and $k$ binning of~\cite{bib:mcdonald06} is adopted.} 
\label{fig:FitRes}
\end{center}
\end{figure}

The purpose of this section is to compare the SDSS and BOSS measured power spectra. We therefore use the same binning  in $k$ and $z$ and the same strategy for the fit as  in \citet{bib:mcdonald06}. In this study, the minimization of $\chi^2(P,\sigma;\Theta)$ was performed with the MINUIT package~\citep{bib:minuit}. Fig.~\ref{fig:FitRes} shows the $P(k_i,z_j)$ measurements of the power spectrum with the $P^{th}(k,z)$ function adjusted by the minimization of $\chi^2(P,\sigma;\Theta)$. The results of the fit are given in Table~\ref{tab:resfit}, and the determination of the error with a frequentist interpretation is discussed in the next paragraph.

\subsection{Frequentist interpretation}

Most recent Ly$\alpha$ analyses use Markov Chain Monte Carlo simulations \citep{bib:viel10} with Bayesian inference. The 
debate between the Bayesian and the frequentist statistical approaches is beyond the scope of this paper.  The philosophical difference between the two methods should not generally lead, in the end, to major differences in  determining  physical parameters and their confidence intervals when the parameters stay in a physical region (see~\cite{bib:yeche06}).

Our work is based on the `frequentist' (or `classical') confidence-level method originally defined by \citet{bib:neyman}.  This avoids any potential bias due to the choice of priors.  In addition, we have also found ways to improve the calculation speed, which gives our program some advantages over  Bayesian programs. 

We first determine the minimum $\chi^2_0$ of $\chi^2(x,\sigma_{x};\Theta)$ leaving  all the cosmological parameters free. Then, to set a confidence level (CL) on any individual cosmological parameter $\theta_i$, we scan the variable $\theta_i$: for each fixed value of $\theta_i$, we again minimize  $\chi^2(x,\sigma_{x};\Theta)$ but with $n-1$ free parameters. The $\chi^2$ difference, $\Delta \chi^2(\theta_i)$, between the new minimum and  $\chi^2_0$, allows us to compute the CL on the variable, assuming that the experimental errors are Gaussian,
\begin{equation}
{\rm CL}(\theta_i) = 1-\int_{\Delta \chi^2(\theta_i)}^{\infty}  f_{\chi^2}(t;N_{dof}) dt,
\label{Eq:CL}
\end{equation}
with
 \begin{equation}
 f_{\chi^2}(t;N_{dof})=\frac{e^{-t/2}t^{N_{dof}/2 -  1}}{\sqrt{2^{N_{dof}}} \Gamma(N_{dof}/2)}   \label{Eq:chi2}
\end{equation}
where $\Gamma$ is the Gamma function and the number of degrees of freedom $N_{dof}$
is equal to 1.
This method can be easily extended to two variables. In this case, the minimizations are
performed for $n-2$ free parameters and the confidence level ${\rm CL}(\theta_i,\theta_j)$ is
derived from Eq.~\ref{Eq:CL} with $N_{dof}=2$.

By definition, this frequentist approach does not require any marginalization to determine the sensitivity on a single individual
cosmological parameter.  Moreover, in contrast to Bayesian treatment, no prior on the cosmological parameters is needed.  With this approach, the correlations between the variables are naturally taken into account and the minimization fit can explore the whole phase space of the cosmological, astrophysics, and nuisance parameters.

\begin{figure}[htbp]
\begin{center}
\epsfig{figure= 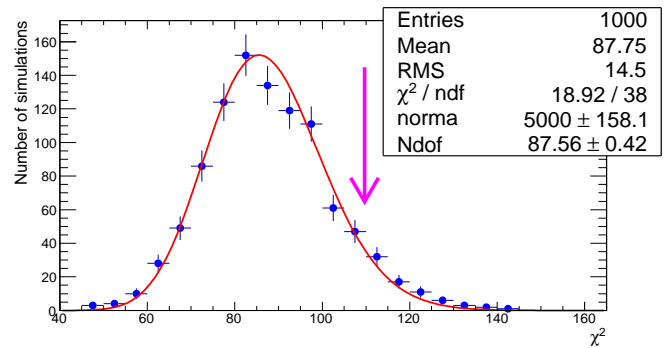,width = \columnwidth} 
\caption[]{\it The distribution of the $\chi^2$ obtained for 1000 simulations reproducing the 1D power spectrum measured by BOSS. The mean of the distribution is 87.8 and a fit of the distribution following a $\chi^2$ law defined in Eq.~\ref{Eq:chi2} gives a number of degrees of freedom equal to $87.6\pm0.4$.} 
\label{fig:GOF}
\end{center}
\end{figure}

It is difficult to define the goodness of fit to the data with the absolute value of $\chi^2_0=112.1$  since we do not know  the actual number of degrees of freedom of our problem because of  correlations between the fit parameters. Therefore, we perform 1000 simulations of the measured power spectrum (96 measurements) and  repeat the fit for each simulation. The distribution of the $\chi^2$ can be used to derive the goodness of fit. A fit of the distribution of Fig.~\ref{fig:GOF} indicates a number of degrees of freedom equal to $87.6\pm0.4$. The fraction of simulations having a $\chi^2$ value higher than  $\chi^2_0$  is six percent.

\subsection{Results on $\sigma_8$ and $n_s$}

\begin{figure}[htbp]
\begin{center}
\epsfig{figure= 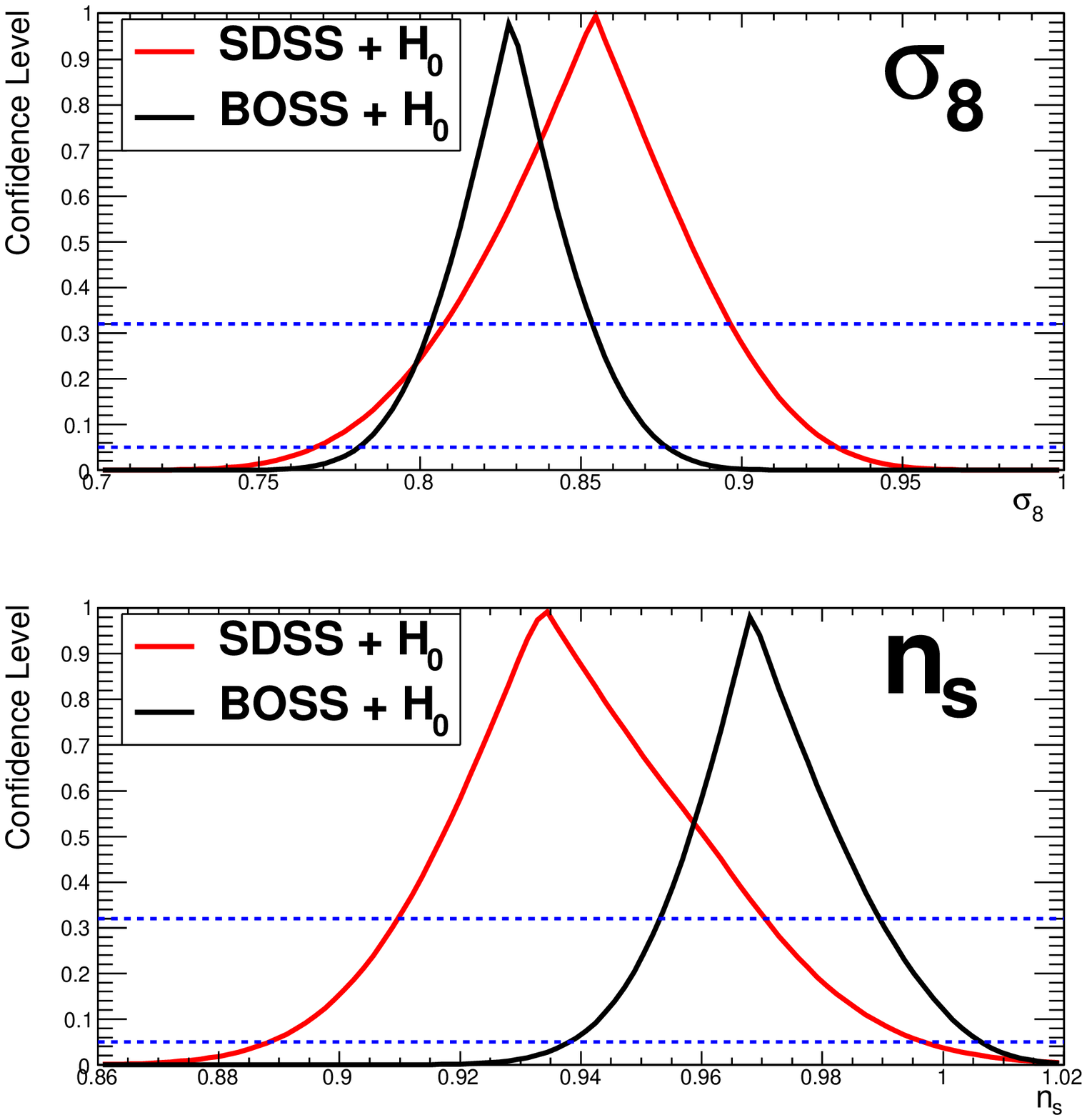,width = \columnwidth} 
\caption[]{\it Confidence level  for the $\sigma_8$ and $n_s$ cosmological parameters with a frequentist interpretation. The red and black curves are obtained respectively for SDSS and BOSS measurements of the power spectrum.} 
\label{fig:Contour1D}
\end{center}
\end{figure}

\begin{figure}[htbp]
\begin{center}
\epsfig{figure= 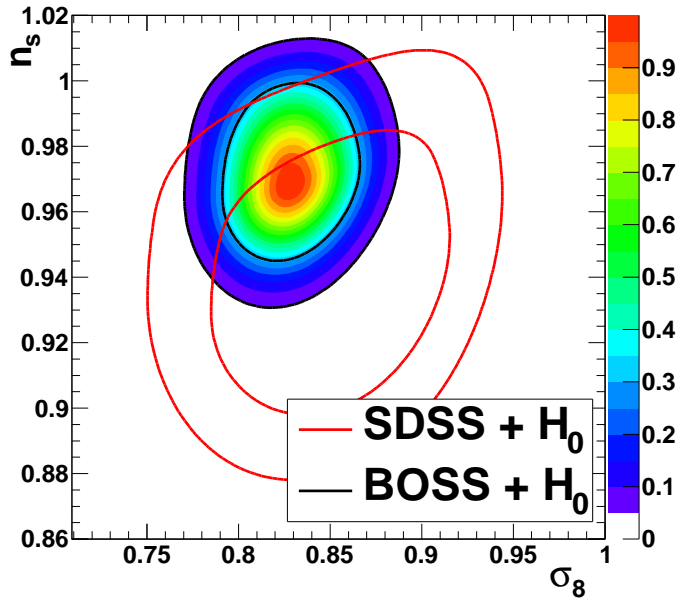,width = \columnwidth} 
\caption[]{\it 2D confidence level contours for  the $\sigma_8$ and $n_s$ cosmological parameters with a frequentist interpretation. The red and black curves are obtained respectively for SDSS and BOSS measurements of the power spectrum.} 
\label{fig:Contour2D}
\end{center}
\end{figure}

Table~\ref{tab:resfit} shows the results for  the cosmological and the astrophysical parameters. Since the hydrodynamical simulations are preliminary versions of simulations that we are currently developing, we  only focus here on the measurement of the two cosmological parameters $\sigma_8$ and $n_s$ in the case of  $\Lambda$CDM, assuming a flat Universe with three neutrinos with $\sum m_\nu = 0$.  In this interpretation section, the other parameters should be considered as nuisance parameters, and the constraints obtained are only indicative. They will be refined in future work.  Figures~\ref{fig:Contour1D} and~\ref{fig:Contour2D} present the confidence level obtained with Eq.~\ref{Eq:CL} on $n_s$ and $\sigma_8$. The central values of the fits on the SDSS measurements of \cite{bib:mcdonald06} and on the BOSS measurements agree within $1\sigma$. The comparison of the confidence level curves shows an improvement by a factor 2-3 in the constraint on these parameters with the BOSS data compared to the SDSS data.  We measure $\sigma_8 =0.83\pm0.03$ and $n_s= 0.97\pm0.02$ in the \ion{H}{i} absorption range  $z= [2.1-3.7]$. 

Since the two measurements of $H_0$ by HST and by Planck  are not fully consistent, we have chosen to use a conservative constraint on $H_0$ that encompasses both and has a constant probability between the two measurements.

In our quasar selection,  we  removed a large portion of the DLA. Therefore in our likelihood we have fixed to zero the amplitude $A_{\rm damp}$  of the component modeling the effect of the DLA in the  power spectrum. If we float this parameter, 
we get  $A_{\rm damp}= 0.14\pm0.10$, which is close to 0 as expected, and the values of $\sigma_8$ and $n_s$ are unchanged. 

The possible systematic uncertainties are included through nuisance parameters in the fit. The errors on $\sigma_8$ and $n_s$ are not dominated by a unique category of parameters. Smaller errors by factor $\sim2$ (respectively $\sim3$)  would be obtained on both parameters if we ignore nuisance parameters (resp. astrophysical parameters). 

\begin{table}[htdp]
\caption{Results of the fit (frequentist approach) to the measured $P(k_i,z_j)$ for the first eight redshift bins  covering the $z= [2.1-3.7]$ region. We used a conservative constraint on $H_0$ that encompasses the measurements given by~\cite{riess11} and by~\cite{bib:planck13}.  }
\begin{center}
\begin{tabular}{lc}
\hline
Parameter & Value \\
\hline
$\sigma_8$ &  $0.83\pm0.03$ \\
$n_s$ &  $0.97\pm0.02$ \\
$\Omega_m$ &  $0.26\pm0.04$ \\
$H_0$ &  $74^{+ 2}_{-7}\; {\rm km/s/Mpc}$ \\
$ \tau_{{\rm eff,}z=3}^A$ &  $0.34\pm0.02$ \\
$ \tau_{\rm eff,}^S$ &  $3.1\pm0.2$ \\
$T^A_{0, z=3}$ &  $(28\pm 5)\times 10^3$ \\
$T^S_{0}$ &  $-3.8\pm1.2$ \\
$\gamma^A_{ z=3}$ &  $0.4\pm0.3$ \\
$\gamma^S$ &  $-6.6\pm3.4$ \\
$f_ {\rm{Si\,III}}$ &  $0.009\pm0.001$ \\
\end{tabular}
\end{center}
\label{tab:resfit}
\end{table}%

\subsection{Discussion and prospects for  future simulations}

The preliminary cosmological results presented in the previous section are  based on a set of hydrodynamical simulations by \citet{bib:viel06}, which are 
sufficient for our purposes but can be improved to meet the more stringent requirements of the BOSS survey. The restriction to only eight redshift bins is dictated by the set of simulations currently available for the analysis. Our next goal is  to use the full redshift information  in the 1D power spectrum, with all twelve redshift bins up to $z=4.5$. To this end, we are following two lines of research, which will be presented in forthcoming publications.
One is to upgrade the quality of the hydrodynamical simulations for the Lyman-$\alpha$ forest, both in resolution (i.e. number of particles)
and box size. 
The other is to include massive neutrinos in our hydrodynamical simulations at the sensitivity of the BOSS survey, along 
with a number of technical and conceptual improvements, with the aim of constraining or measuring the sum of  neutrino masses.

%%%%%%

\section{Conclusions}\label{sec:discussion}

We have developed two independent methods for measuring the 1D power spectrum of the transmitted flux in the Ly$\alpha$  forest. The first method is based on a Fourier transform, and  the second approach relies upon a maximum likelihood estimator. The two methods are independent and present different systematic uncertainties owing to the  techniques used to  mask  pixels contaminated by sky emission lines or to take  the spectrograph resolution and the noise contribution to the Ly$\alpha$ power spectrum into account, which differ in the two approaches.  Determining the noise level in the data spectra was subject to a novel  treatment, because of its significant impact  on the derived power spectrum. 

 We applied these two methods to 13~821 quasar spectra from SDSS-III/BOSS, that were  selected from a larger sample of almost 90~000 DR9 BOSS spectra on the basis of their high quality,  high S/N, and  the low value of the spectral resolution. The  power spectra  measured using either method  are in good agreement over all twelve redshift bins from $\langle z\rangle = 2.2$ to $\langle z\rangle = 4.4$. We determined the systematic uncertainties on our measurements coming both from the analysis  method and from our knowledge of the instrument characteristics.

We presented a preliminary cosmological interpretation of our experimental results, 
along the lines of~\citet{bib:viel06}, limiting the analysis to the first eight redshift bins up to $z=3.7$. The improvement in precision over  previous studies from SDSS~\citep{bib:mcdonald06} allows for a factor 2--3 tighter constraints on relevant cosmological parameters. In particular, for a $\Lambda$CDM model and using a constraint on $H_0$ encompassing measurements from  the HST and from Planck, we measure $\sigma_8 =0.83\pm0.03$ and $n_s= 0.97\pm0.02$ in the \ion{H}{i} absorption range  $2.1<z<3.7$. 

These results were obtained by assuming a flat $\Lambda$CDM model with no massive neutrinos.  In the near future, we will update the cosmological interpretation of our results by using the full information contained in the 1D power spectrum over all twelve redshift bins up to $z=4.5$. This will be done thanks a new set of hydrodynamical simulations for the Ly$\alpha$ forest that we will run with an upgrade in both resolution and box size to match the sensitivity of our measurement and, in addition, which will include massive neutrinos.

\begin{acknowledgements}
Funding for SDSS-III has been provided by the Alfred P. Sloan Foundation, the Participating Institutions, the National Science Foundation, and the U.S. Department of Energy. The
SDSS-III web site is http://www.sdss3.org/.\\
The BOSS French Participation Group is supported by Agence Nationale de la Recherche under grant ANR-08-BLAN-0222. A.B., N.P.-D., G.R., and Ch.Y.  acknowledge  support from grant ANR-11-JS04-011-01 of Agence Nationale de la Recherche. MV is supported by ERC-StG ``CosmoIGM".\\
SDSS-III is managed by the Astrophysical Research Consortium for the Participating Institutions of the SDSS-III Collaboration including the University of Arizona, the Brazilian Participation Group, Brookhaven National Laboratory, University of Cambridge, University of Florida, the French Participation Group, the German Participation Group, the Instituto de Astrofisica de Canarias, the Michigan State/Notre Dame/JINA Participation Group, Johns Hopkins University, Lawrence Berkeley National Laboratory, Max Planck Institute for Astrophysics, New Mexico State University, New York University, the Ohio State University, the Penn State University, University of Portsmouth, Princeton University, University of Tokyo, the University of Utah, Vanderbilt University, University of Virginia, University of Washington, and Yale University.
\end{acknowledgements}

\bibliographystyle{aa}
\bibliography{biblio}

\end{document}